\documentclass[10pt,letterpaper]{article}
\usepackage[top=0.85in,left=1.4in,footskip=0.75in]{geometry}

\usepackage{xcolor}
\usepackage{amsmath,amssymb}

\usepackage{changepage}

\usepackage[utf8x]{inputenc}

\usepackage{textcomp,marvosym}
\usepackage[export]{adjustbox}

\usepackage[superscript]{cite}

\usepackage{xcolor}
\usepackage{nameref}

\usepackage[unicode=true,pdfusetitle,
bookmarks=true,bookmarksnumbered=false,bookmarksopen=false,
 breaklinks=false,pdfborder={0 0 0},pdfborderst yle={},backref=false,colorlinks=true]
 {hyperref}
 \hypersetup{
 citecolor=blue,
 urlcolor=teal,
 linkcolor=cyan}

\usepackage{microtype}

\usepackage{cellspace}
\usepackage{array}
\usepackage{svg}
\newcolumntype{+}{!{\vrule width 2pt}}

\newlength\savedwidth

\raggedright
\usepackage[aboveskip=1pt,labelfont=bf,labelsep=period,justification=raggedright,singlelinecheck=off]{caption}

\usepackage{tabularx}
\newcolumntype{Y}{>{\centering\arraybackslash}X}

\usepackage[final]{pdfpages}

\usepackage{physics}
\usepackage{float}

\newcommand{\cS}{\mathcal{S}}

\usepackage[section]{placeins}

\usepackage[table]{}
\usepackage{array}
\usepackage{multirow}
\usepackage{multicol}
\usepackage{color, colortbl}
\definecolor{BlueGreenish}{rgb}{0.37,0.63,0.62}
\definecolor{Orange}{rgb}{0.93,0.53,0.18}
\definecolor{Green}{rgb}{0,0.53,0}
\begin{document}

\makeatletter
\def\Cline#1#2{\@Cline#1#2\@nil}
\def\@Cline#1-#2#3\@nil{%
  \omit
  \@multicnt#1%
  \advance\@multispan\m@ne
  \ifnum\@multicnt=\@ne\@firstofone{&\omit}\fi
  \@multicnt#2%
  \advance\@multicnt-#1%
  \advance\@multispan\@ne
  \leaders\hrule\@height#3\hfill
  \cr}
\makeatother

\vspace*{0.1in}

\begin{flushleft}
  {\Large 
  
  

   Symmetries and synchronization from whole-neural activity in {\it C. elegans} connectome: Integration of functional and structural networks}
 \end{flushleft} 

 


Bryant Avila\textsuperscript{1*},
Pedro Augusto\textsuperscript{2,3*},
David Phillips \textsuperscript{4},
Tommaso Gili \textsuperscript{5},
Manuel Zimmer \textsuperscript{2$\dagger$},
Hern\'an~A.~Makse\textsuperscript{1,6,7$\dagger$}

\bigskip
\textbf{1} Levich Institute and Physics Department, City College  of New York, New York, NY 10031, USA

\textbf{2} Department of Neuroscience and Developmental Biology, University of Vienna, Vienna Biocenter (VBC), Vienna, Austria

\textbf{3} Vienna Biocenter PhD Program, Doctoral School of the University of Vienna and Medical University of Vienna, Vienna, Austria

\textbf{4} Mechanical Engineering Department, University of New Mexico, Albuquerque, NM
87131, USA

\textbf{5} Networks Unit, IMT Scuola Alti Studi Lucca, Piazza San Francesco 15, 55100, Lucca, Italy

\textbf{6} Department of Radiology, Neuroradiology Service, Memorial Sloan Kettering Cancer Center, New York, NY 10065, USA

\textbf{7} CUNY Neuroscience, Graduate Center, City University of New
York, New York, NY 10031, USA
\bigskip

\textsuperscript{*} Equal contribution 
\textsuperscript{$\dagger$} Corresponding author, manuel.zimmer@univie.ac.at  
\textsuperscript{$\dagger$}Corresponding author, hmakse@ccny.cuny.edu


\section*{Abstract}
Understanding the dynamical behavior of complex systems from their
underlying network architectures is a long-standing question in
complexity theory. Therefore, many metrics have been devised to
extract network features like motifs, centrality, and modularity
measures. It has previously been proposed that network symmetries are
of particular importance since they are expected to underly the
synchronization of a system's units, which is ubiquitously observed in
nervous system activity patterns. However, perfectly symmetrical
structures are difficult to assess in noisy measurements of biological
systems, like neuronal connectomes. Here, we devise a principled
method to infer network symmetries from combined connectome and
neuronal activity data.  Using nervous system-wide population activity
recordings of the \textit{C.elegans} backward locomotor system, we
infer structures in the connectome called fibration symmetries, which
can explain which group of neurons synchronize their activity. Our
analysis suggests functional building blocks in the animal's motor
periphery, providing new testable hypotheses on how descending
interneuron circuits communicate with the motor periphery to control behavior.  Our approach opens a new door
to exploring the structure-function relations in other complex
systems, like the nervous systems of larger animals.

\section*{Significance Statement}
Complex biological networks exhibit an intricate relationship between the structural composition of their elements’ interaction and the whole system's functionality. This is particularly true in neuroscience, where the multiscale organization of neurons gives place to a multitude of hierarchical functionalities responsible for the nervous system interacting with the environment in which it is embedded. In this study, we focus on the chemical synaptic network of the backwards-crawling locomotion circuitry of the worm C. elegans to correlate structural clusters of neurons with dynamic patterns of their synchronization. Our findings reveal functional building blocks in the animal's motor periphery, providing new testable hypotheses on how descending interneuron circuits communicate with the motor periphery to control behavior.

\section{Introduction} \label{sec:intro}
Complex systems depend on an intricate interplay between the dynamical
properties of their building blocks and the network structures that
link them together. For decades, researchers have endeavored to
uncover how the topological attributes of a network could shed light
on its dynamical behavior. While several methods have been proposed to
elucidate this interplay
\cite{sporns2013structure,friston2011functional,park2013structural,yan2017network},  
the role of the network's symmetries has garnered considerable
attention in recent years
\cite{pecora2014cluster,sorrentino2016complete,stewart2003symmetry,golubitsky2006nonlinear,aguiar2018synchronization,boldi2002fibrations,morone2019symmetry,morone2020fibration}. Identifying
and understanding such symmetries has potentially profound
implications: network theory implies that certain symmetrical
structures termed fibration symmetries in a graph permit
synchronizations
\cite{stewart2003symmetry,golubitsky2006nonlinear,boldi2002fibrations,morone2020fibration},
a ubiquitously observed phenomenon in brain networks
\cite{buzsaki2006rhythms}. Thus, identifying these symmetries might be
crucial for decoding some of the structure-function relationships in a
neuronal network.

\smallskip
Fibration symmetry refers to a specific type of symmetry where nodes
can be grouped into equivalence classes of balanced colorings based on
their input patterns
\cite{boldi2002fibrations,morone2020fibration}. Nodes in the same
class receives identical 'input trees', which are hierarchical mappings
of input connectivity patterns, are proposed to share the same dynamics, and thus can
synchronize their behavior. These equivalent nodes are said to belong
to the same 'fiber' and are also 'balanced colorings' of the graph
\cite{golubitsky2006nonlinear}. This is formalized through admissible
ordinary differential equations (ODEs), which describe the evolution
of the system's state \cite{pecora2014cluster}. Theory ensures that
balanced colored nodes with isomorphic input trees following a set of
admissible ODEs will evolve in synchrony
\cite{stewart2003symmetry,golubitsky2006nonlinear,deville2015modular,morone2020fibration,leifer2020circuits}
forming a cluster of synchronization
\cite{pecora2014cluster,sorrentino2016complete}. This synchronization
can be crucial in various biological systems ranging from gene
regulatory networks to brain networks involved in language processing
\cite{leifer2021predicting,gili2024fibration}. Thus, identifying
fibration symmetries provide a prediction in which subsets of nodes
potentially exhibit synchronized dynamics.

\smallskip
However, there are technical and conceptual challenges to this
theory. Biological networks like connectomes are inherently
incomplete, and measuring them is limited by annotation errors, 
missing links, and noise, yet identifying fiber symmetries is sensitive
to small variations in a connectivity matrix. More fundamentally,
assuming mathematically perfect symmetries and equivalences of nodes
in a biological network is unrealistic. Here, we propose that network
symmetries represent constraints rather than perfect blueprints for
their implementations in individual animals. We aim to address these
challenges by studying the interplay of symmetry and synchronization
in the locomotor system of the nematode \textit{C.elegans}.

The nematode worm \textit{C.elegans} \cite{sulston1977post,sulston1983embryonic,taylor2021molecular}exhibits a small nervous system
of just 302 neurons that develop from a stereotypic cell lineage into
118 anatomically and genetically defined cell classes, most of which
are comprised of a bilaterally symmetrical pair of neurons (e.g., AVAL
and AVAR) \cite{white1986structure}.  Its connectome has been fully
reconstructed with a synaptic resolution
\cite{white1986structure,chen2006wiring,cook2019whole}, and it is
tractable for large-scale single-cell resolution neuronal calcium
imaging \cite{kato2015global,Uzel2022neuronhubs}.  Both, in
combination, offer unique opportunities to study structure-function
relationships in the nervous system \cite{yan2017network}.  The worm connectome shares many
features with connectivity data from larger nervous systems, e.g.,
rich-club architecture and modularity 
\cite{varshney2011structural,cook2019whole,witvliet2021connectomes,towlson2013rich}. 
Moreover,
neuronal recordings in \textit{C. elegans} revealed nervous
system-wide neuronal activity dynamics that exhibit synchronization
patterns among well-defined ensembles of
neurons\cite{kato2015global,Uzel2022neuronhubs,kaplan2020nested}. These
dynamics correspond to motor commands for a set of actions like
forward-crawling, backward-crawling, or
turning\cite{kato2015global,atanas2023brain}.

These dynamics are generated in the absence of any known acute
time-varying sensory stimuli
\cite{kato2015global,Uzel2022neuronhubs,kaplan2020nested} and can be
promoted by arousing conditions, such as environmental oxygen or blue
light \cite{nichols2017global, gauthey2024light}. Moreover, these
dynamics can be observed in immobilized animals; even under such
conditions, they extend to the motor periphery, incorporating motor
neurons, under unconstrained conditions mediating movement execution
\cite{kaplan2020nested}. These features point toward intrinsic
mechanisms that drive and maintain synchronous activity states. In
accordance, our previous work indicated that rich club architecture
and input similarities are crucial architectural features in the
connectome that permits synchronized neuronal
dynamics\cite{Uzel2022neuronhubs}.

In the present study, we focus on the chemical synaptic network of the
backward-crawling locomotion circuitry of \textit{C. elegans} to circumvent some of the technical and conceptual constraints posed
by theory and computational limits.  This sub-network of just 21 individual
neurons includes three bilateral pairs of interneurons, AVA, AVE, and AVD,
which are situated in the head of the animal and send descending
chemical synapses to two classes of motor neurons, termed DA1-9 and
VA1-12. These motorneurons form neuromuscular junctions with the
dorsal and ventral body wall muscles, respectively, and are required
to generate the backward crawling gait \cite{chalfie1985neural, kawano2011imbalancing,gao2018excitatory}. All
chemical synapses in this sub-network are cholinergic and, with few
exceptions, neurons within the same class share similar morphology and
gene expression patterns \cite{white1986structure, pereira2015cellular,taylor2021molecular}. Hence, assuming equivalence among nodes within
the same class could be a reasonable simplification.

By leveraging advanced calcium imaging
techniques, we first study neuronal dynamics, aiming to identify
patterns of synchronization within its backward locomotion circuitry. We
then use the observed synchronization dynamics to infer underlying symmetries in the
connectome.

Given the potential variability in connectomes across individual worms,
expecting the universal connectome model to predict the exact neural
synchronization across neurons is unrealistic. This variability
underscores the necessity of adopting tailored approaches to
understand the dynamic behavior from the structure of the
connectome. We approach this problem by reconstructing the underlying
connectome of {\it C. elegans} guided by the synchronization dynamics
obtained experimentally. This is done by "repairing" the connectivity
structure of a typical connectome \cite{varshney2011structural} by
solving an appropriately modeled integer linear program to
find the minimal number of repairs of the available connectome needed
to ensure the observed synchronization dynamics. We aim for a
fibration-symmetric connectome that reflects the observed
synchronization. This method provides a means to idealize connectomes conceptually.  According to inter-animal connectome differences reported experimentally in
\cite{witvliet2021connectomes,hall1991posterior}, 
the threshold for connectome modifications can be taken to not exceed approximately 50\%. However, for our final repair solution we take the stringent threshold of 25\% given by animal-to-animal differences found in \cite{varshney2011structural}.

To ascertain the robustness and validity of our solutions, we engage
in a rigorous testing procedure. We aim to determine the optimality of
our solution by reshuffling the neuronal labels in the connectome and
revisiting the repair process.  We accept our method's solution only
if the post-relabeling solutions consistently underperform our primary
findings (in over 95\% of instances).

Our study bridges the gap between structural and functional neuronal
data, leveraging the power of symmetries in graph theory to shed light
on synchronization dynamics within \textit{C. elegans} and the
structure of the connectome underlying this synchronization.


\section{Results} \label{theory}

\begin{figure}[H]
  \centering
  \includegraphics[width=\linewidth]{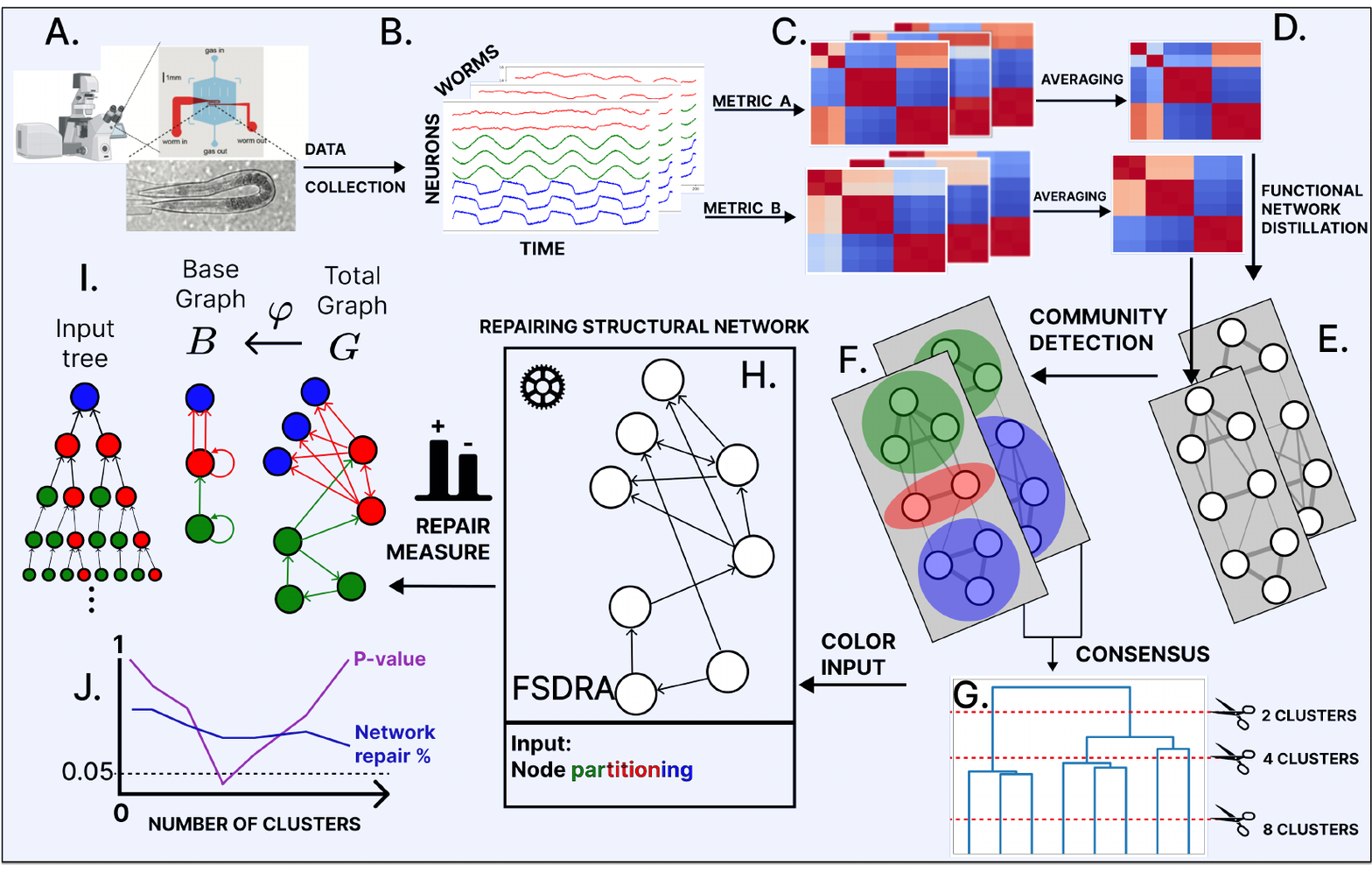}
  \captionsetup{font=small}
\caption{\textbf{Pipeline for the structural network repair of the backward locomotion of {\it C. elegans} based on neural recordings.}
(A) Experimental setup.
(B) Time series data showing the activity traces of multiple neurons. Traces are color-coded to indicate similar dynamics.
(C) Matrices of synchronization are obtained from the traces.
(D) The synchronicity matrices are averaged across worms, accounting for any missing activity traces of individual neurons.
(E) The percolation procedure was applied to the averaged matrices to obtain the functional network.
(F) Averaging the functional networks leads to a consensus matrix across different methods of synchrony measured. 
(G) A hierarchical partitioning is implemented to identify clusters of synchronicity.
(H) MILP repairs the network according to cluster synchronization.
(I) The solution to the MILP produces an ideal network guided by synchronization. This network can be "collapsed" into its base.
(J)  p-value statistics to choose the optimal solution. 
}
\label{fig:pipeline}
\end{figure}

In this section, we delineate the pipeline employed throughout our
study in a condensed fashion and details are elaborated in the
Methods section \ref{Methods}.  Figure \ref{fig:pipeline} provides an
outline of the method to reconstruct the connectome:

\begin{enumerate}
\item The
process begins with the experimental setup to record neuron activity in  Fig. \ref{fig:pipeline}A with a microfluidic device
for immobilization of {\it C. elegans} enabling nervous system-wide Ca++ imaging with confocal
fluorescence microscope setup.

\item Time series data 
for activity traces of multiple neurons are recorded simultaneously, as
 shown in Fig. \ref{fig:pipeline}B. This
procedure is performed on multiple worms under similar
conditions. 

\item To obtain functional pairwise synchrony matrices, various metrics that capture synchrony are applied to the
  recorded time series data(Fig. \ref{fig:pipeline}C).

\item The synchrony matrices shown in Fig. \ref{fig:pipeline}D are
  averaged across worms to account for trial-to-trial and/or inter-individual variability. 

\item A standard thresholding process is then applied to obtain the
  functional network (Fig. \ref{fig:pipeline}E)
  \cite{rubinov2010complex,gallos2012small}. Starting from a
  disconnected graph, we add links between nodes in decreasing order
  of weight of the averaged synchrony matrix. This is done for each
  averaged matrix.

\item Finally, a consensus matrix is calculated
  \cite{tian2022scmelody} across different methods of synchrony
  (Fig. \ref{fig:pipeline}F).

\item A hierarchical clustering algorithm is implemented to find a
  partition of synchrony clusters (Fig. \ref{fig:pipeline}G). Each
  neuron is assigned a color according to its cluster of
  synchrony. These colors are used as inputs to the repair algorithm
  in the next step.

\item We formulate and solve a mixed integer linear program (MILP)
  that finds the minimum number of edges to add or remove from the
  raw connectome (taken from
  \cite{varshney2011structural}) to produce an ideal
  fibration-symmetric network that reproduces the coloring in the
  consensus partition obtained in the previous step
  (Fig. \ref{fig:pipeline}H).

\item The network produced by the solution to the MILP is a fibration symmetric connectome with cluster synchronization that reproduces the experimental data. This network can be collapsed into a smaller representation, the base
  graph, where nodes belonging to the same fiber have isomorphic input
  trees (Fig. \ref{fig:pipeline}I). The base graph suggests functional building blocks in the connectome.

\item For each consensus
  partition, a permutation p-value test is performed by permuting
  node labels and repairing the structural network 1,000 times. The partition with the lowest p-value is
  chosen as the optimal solution (Fig. \ref{fig:pipeline}J).
\end{enumerate}


\subsection{Neuronal activity recordings} \label{sec:data_collection_setup}

To simultaneously record calcium activity from interneurons
in the head and motor neurons along the entire ventral cord, we
modify a recently reported whole nervous system imaging pipeline
\cite{kaplan2020nested}. Briefly, worms expressing the calcium probe
NLS-GCaMP6f in a pan-neuronal fashion, together with NeuroPal
multi-color cell identification labels \cite{yemini2021neuropal}, are
immobilized in a microfluidic device that positions them in the field
of view of a spinning disk confocal microscope. After fast volumetric
GCaMP6f imaging, multi-color stacks are taken for cell class
identification (see Methods section 
\ref{sec:apparatus_and_setups} for details).

\subsection{Whole nervous system recording and characterization of motor neuron activity} \label{ca2plus}

\begin{figure}[H]
  \centering
    \includegraphics[width=0.75\linewidth]{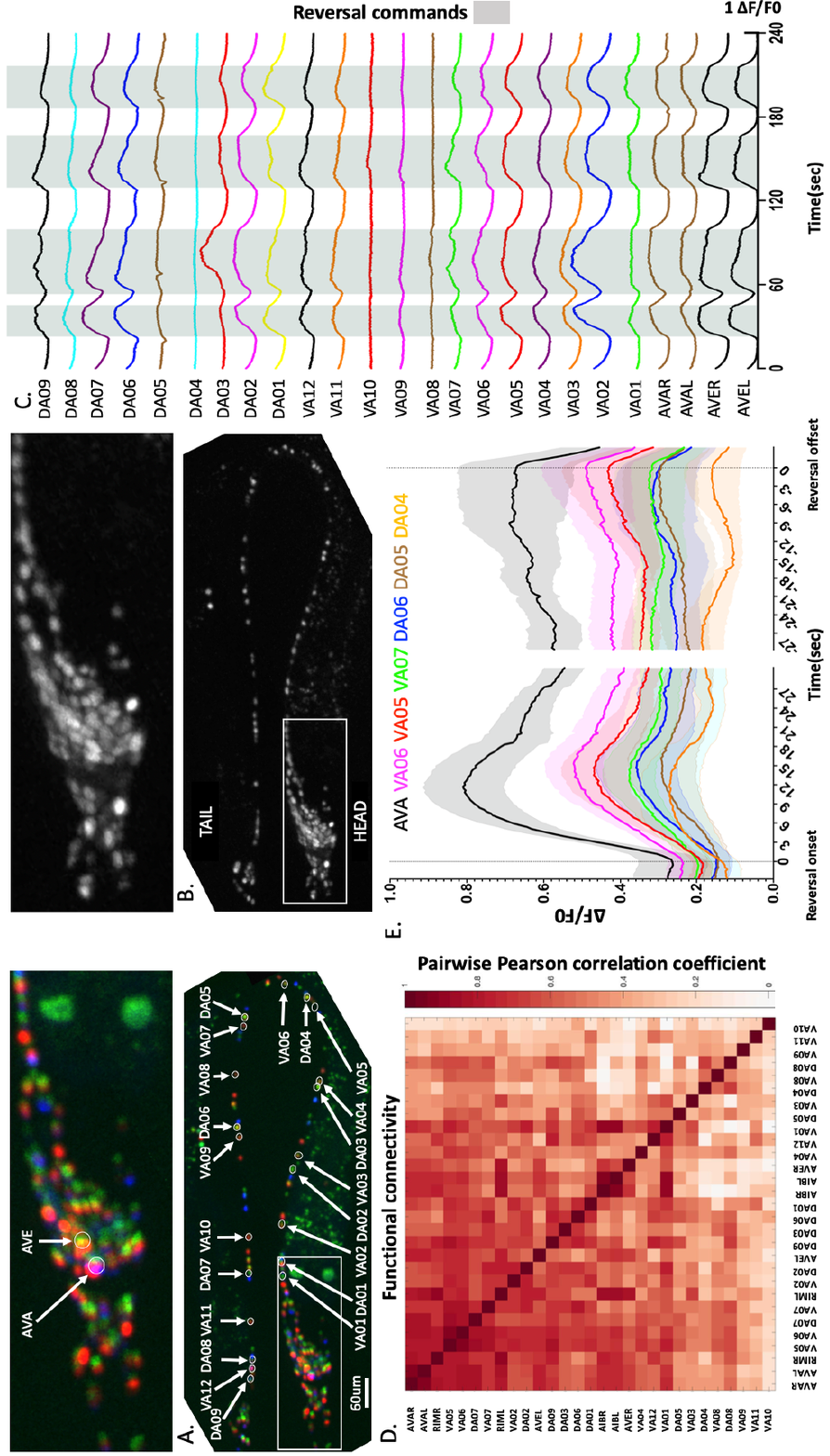}
\caption{{\bf Whole nervous system recording reveals synchronous motor
    neuron activity}. (A) NeuroPAL labeled worm with selected neuronal
  cell class identities indicated. (B) The same worm as in (A) showing
  NLS-GCaMP6f labeling. (C) Activity time series (DF/F0) of selected,
  grey shading indicates reversal command states defined by AVAL
  activity.  (D) Pairwise Level of Synchronization ($LoS$) among
  selected neurons. Each matrix entry is the average of $N=3-8$
  pairwise observations.  (E) The triggered average ($\pm$SEM) to
  reversal command onset (left) and offset (right) are both defined by
  reference neuron AVAL. Three example neurons of each VA and DA motor
  neuron class are shown.  Averages calculated across recordings
  represent the dynamics that we observe in our datasets; neurons are
  relatively aligned in time, but there are substantial differences
  regarding their amplitudes.}
\label{fig:neuroPAL}
\end{figure}

We generated eight datasets from different young adults well-fed
worms. Imaging covered almost the entire body spanning from the head
ganglia, the complete ventral cord, and the tail ganglia with
single-cell resolution (Fig. \ref{fig:neuroPAL}A-B). Animals were
recorded for 10 minutes at approximately three volumes per
second. Figure \ref{fig:neuroPAL}C shows a multi-neuron time series
with discernible calcium activity patterns of the neurons selected for
this study. These include neurons AVAL/R and AVEL/R, which belong to
the major descending interneurons conveying motor commands for
backward crawling, as well as the downstream backward crawling motor
neurons of the DA and VA classes. Consistent with previous studies
\cite{kato2015global,kaplan2020nested,Uzel2022neuronhubs}, AVA and AVE
activity exhibited discrete transitions in their activity patterns,
characterized as low, rise, high, and fall states. We previously
validated that these states correspond to behavioral states in freely
crawling animals, i.e., low corresponds to forward crawling, rise and
high corresponds to backward crawling, and fall corresponds to turning
\cite{kato2015global}.

Here, we observe that these features are mirrored in the motor
periphery (Fig. \ref{fig:neuroPAL}C). The mean pairwise correlations
among motorneurons and interneurons were high, indicating the strong
coupling between them (Fig. \ref{fig:neuroPAL}D). During free backward
crawling, animals generate a posterior-to-anterior traveling body wave
of alternating dorsal-ventral body oscillations. We did not observe
any obvious oscillations or activity patterns alternating between the
DA and VA motorneurons, which innervate the dorsal and ventral body
wall muscles, respectively. Neither did we observe any delay or
sequential activation patterns between motor neurons that innervate
more posterior to anterior muscles (which is indicated by the index
number in their name, $n=1$ most anterior, $n=12$ most
posterior). However, all motor neurons follow the rise and fall states
of the AVA descending command interneuron
(Fig. \ref{fig:neuroPAL}E). In conclusion, under our experimental
immobilization conditions, A-class motorneurons do not generate
gait-related patterns in their Ca++ activity profiles. However, they appear to respond to descending
motor commands from their presynaptic interneurons reliably. This
feature and the absence of movement, hence the lack of proprioceptive
inputs, makes the present setup particularly suitable for studying
neuronal dynamics dependent on internal neuronal circuit interactions
(see also discussion).

\subsection{Synchronization measures} \label{sec:synchronization_sea}

Assessing the synchronization of two or more signals recorded
simultaneously involves various methods tailored to the specific
characteristics and similarities of interest. This results in a
diverse array of techniques aiming to capture synchronization through
different approaches \cite{kreuz2013synchronization}. These techniques
are generally classified into four primary groups based on their
focus: time domain versus frequency domain and methods that account
for directional dependencies versus those that do not (see Methods
section \ref{sec:synchronization_measures} and  
Table
\ref{table:synchronization_measures}).


\smallskip
Correlation is a robust
measurement for coherence in neuronal data obtained via calcium
imaging
\cite{kato2015global,hallinen2021decoding,creamer2022correcting}
(Methods section \ref{sec:synchronization_measures}).
In addition to correlation, to quantify the amount of synchronicity
between two signals, we implement the Level of Synchronicity ($LoS$)
measure introduced in \cite{phillips2011gentle,avila2024fibration}.
This metric determines how closely two
signals are to perfect synchronicity, meaning that
two signals have the same value at the same time, forming a cluster of synchrony:

\begin{equation}\label{eq:cluster_synchronicity}
  V_{i}(t) = V_{j}(t)  \,\,\,\,\,\,\,\,\, \forall i,j\in C_k.
\end{equation}
Here, $V_i$ is the dynamic of a neuron, which, in the present case, is
associated with the neuron's calcium activity. $C_k$ is one of the
non-overlapping sets of neurons partitioning the neural system into
\textit{Cluster of Synchronicity} (CS) \cite{pecora2014cluster}.

\smallskip
To quantitatively capture this synchronicity, the $LoS$ evaluates the 
synchronization of two signals over time by considering their instantaneous differences 
and scaling them through a parameter $\sigma$ defined as:

\begin{equation}\label{eq:LoS}
        LoS_{ij} = \frac{1}{T}\sum_{t}^{T} \exp{-\frac{[V_{i}(t) - V_{j}(t)]^2}{2\sigma^2}},
\end{equation}
where $T$ is the total amount of time steps in which
$LoS$ is measured between the signals of neurons $i$ and $j$.
The parameter $\sigma$ serves as a scale to define a benchmark for
closeness between two points in time.

\smallskip
This parameter permits the user to deal with natural variations in
biological systems. No two neurons are perfect copies of each other;
therefore, if identical signals stimulate two similar neurons at rest,
their outputs (membrane potential) will naturally vary in intensity
and phase by small amounts. With this concept in mind, we implement
multiple versions of the $LoS$, each with a different value for
$\sigma$. When the value of $\sigma$ is zero, each signal will only be
synchronous with its exact copy; as the value of $\sigma$ increments,
it reaches the state that all signals are synchronous with all
others. This indicates that an ideal value lies between zero and an
upper limit (the largest difference between signals at some time $t$
should suffice). After many trials, we settle for values ranging from
0.01 up to 0.20
to explore different levels of synchronicity. We proceed with two
classes of synchronization metrics, correlations and $LoS$, each having
variations through different parameters or relational measurements,
making for a total of 44 metrics.

\smallskip

Matrices of synchronization and correlation are computed for each of
the $N=8$ worms under study, capturing the synchronous activities
observed within each individual as depicted in
Fig. \ref{fig:pipeline}C. To obtain a representative overview of
synchronicity across the entire set of worms, these individual
matrices are averaged per type, i.e. for each value of $\sigma$
(Fig. \ref{fig:pipeline}D). However, compiling these averaged matrices
require care, particularly in instances where the activity of
individual neurons are missing for specific worms. Therefore, we
calculate the average by averaging over worms for a particular metric
type and divide each element (neuron-pair functional measurement) in
the summed matrix by the number of times they appear together across
the whole cohort of worms.

\subsection{Identifying  cluster synchronization from the functional network} \label{sec:functional_clusters}

Each element in the averaged matrix contains information on a
neuron pair synchronous relationships used to construct
functional networks composed of all the neurons for the backward
locomotion gait. 
We follow standard thresholding procedures
\cite{rubinov2010complex,gallos2012small,gili2024fibration} to build
the functional network.  Through this method, we purge the functional
data to only contain the strongest links leading to a functional
network, as seen in Fig. \ref{fig:pipeline}E.  Using this network, we
identify neuron cliques that are synchronized via community and
cluster detection algorithms.

A group of neurons that are more synchronous to each other
than other neurons belonging to different groups is considered
equivalent to a CS of neurons. We apply cluster and community detection algorithms to extract these functionally synchronous clusters of neurons from the functional network using two methods:

\begin{itemize}
  \item Clique synchronization \cite{gili2024fibration}: It decides if a node belongs to a synchronous clique if the average value of the edges of the functional network inside the clique is bigger than any of
    its outside edges (see Methods section \ref{sec:synchronization_used}). 
  \item Louvain community detection \cite{blondel2008fast}: It decides the number of clusters and the number of
    nodes in each cluster by optimizing the modularity function as given
    by Eq. \eqref{eq:modularity} in the Methods section \ref{sec:synchronization_used}, which is solely cluster dependent. 
\end{itemize}

\begin{figure}[p]
    \centering
    \includegraphics[width=.8\linewidth]{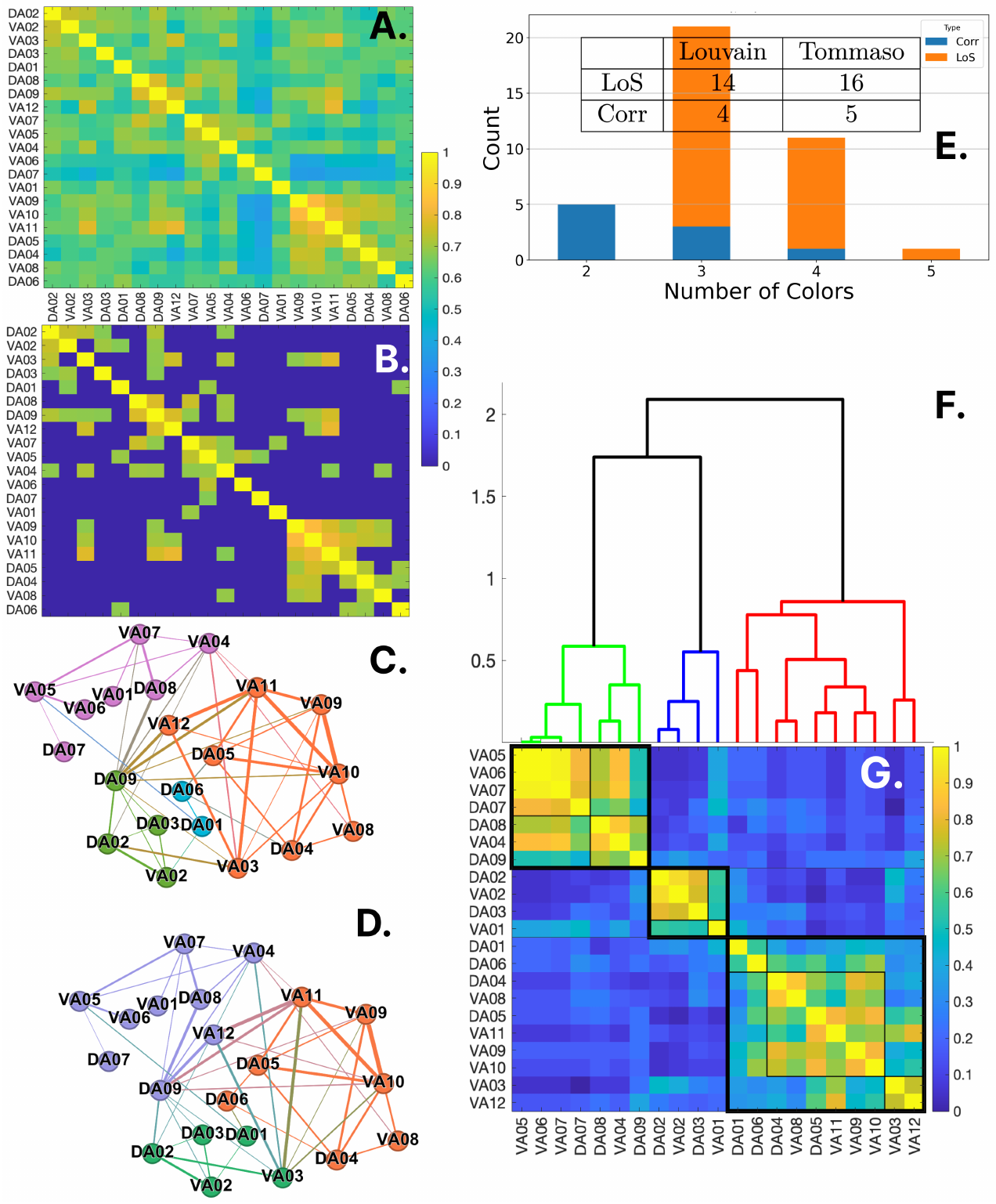}
    \captionsetup{font=small}
      \caption{ \textbf{Results of the clustering analysis.}  (A) Average
        of $N=8$ $LoS$ matrices at $\sigma=0.16$.  (B) Functional matrix
        thresholding removes elements smaller than the smallest and
        largest values for all neurons.  (C) The functional network
        is converted into an indirect graph using the percolation
        thresholding method. Node colors show clustering using the
        Clique Synchronization method. Edge thickness relates to
        values in the $LoS$ matrix.  (D) Same as the previous
        functional network but with the Louvain method partitioning nodes.
        (E) Distribution of synchrony cluster numbers among unique
        partitionings. $LoS$ measurements lead to partitionings with
        more clusters compared to correlation measurements. $LoS$
        peaks at partitionings with 3 clusters; correlations peak at 2
        clusters. Given the neurons' high monotonic and synchronous
        relationships, $LoS$ measurement helps distinguish
        partitionings with more clusters via Ca2+ signal
        amplitude. Partitionings are applied to 39 $LoS$ matrices and
        five correlation matrices using Distance, Pearson, Kendall,
        Spearman, and Covariance. $LoS$ row matrices are obtained from
        unique $\sigma$ values ranging from 0.01 to 0.20 in 0.005
        steps.  (F) Dendrogram of averaged co-occurrence matrices
        using Ward-metric hierarchical clustering. Three major groups
        of backward motor neurons are observed in green, blue, and
        red.  (G) Averaged co-occurrence matrices show two levels of
        hierarchical clustering. Diagonal blocks with thick lines
        correspond to colored neuron groups from the dendrogram, while
        those with thin lines are sub-clusters within the three major
        clusters. These clusters correspond to a dendrogram slice at
        1.00.  }
    \label{fig:multiple_results}
\end{figure}

The result of applying these methods to a $LoS$ correlation matrix
is seen in Figs. \ref{fig:multiple_results}A-B, which results in a functional network with the node partition as seen in Fig. \ref{fig:multiple_results}C.
The nodes with the same colors belong to the same cluster synchronization, as can be seen since they share thicker edges than
edges between nodes of different colors.

The clusters obtained using the $LoS$ measure at
different parametric values produce various motor-neuron partitionings. We only keep those that are unique among
the 39 results for the range of $\sigma$ between 0.01 and 0.20. This
is done for each of the two cluster/community detection methods. The
number of unique partitionings for each method can be seen in the table
within Fig. \ref{fig:multiple_results}E.  

\subsection{Consensus synchronous clusters} \label{sec:consensus_agreement}

We construct 44 metrics to obtain synchronization information: 39
$LoS$ matrices and five types of correlation measures (Pearson,
Spearman, and Kendall coefficient, distance correlation and
covariance). We combine these metrics with the two functional
clustering methods mentioned in the previous section
\ref{sec:functional_clusters} (clique synchronization and Louvain) to
create 88 possible partitionings. Then, a consensus is created
following the methods developed in \cite{tian2022scmelody} to leverage
the information obtained from all the partitionings. To achieve this,
each partitioning, which consists of $n$ clusters of synchronous
neurons, is used to create a co-occurrence matrix, which is simply a
matrix with $n$ diagonal blocks, one for each synchronous group, with
value 1, and with all other values outside these blocks set to zero
(see Fig. \ref{fig:multiple_results}D).  These matrices are summed up and
normalized.  The resulting consensus matrix is depicted in
Fig. \ref{fig:multiple_results}G.  Through this, the sets of neurons
that appear in the same synchronous group more frequently than with
other neurons appear with higher values. This consensus avoids putting
each partitioning in competition with the other and instead compiles
them into a globally agreed-upon result \cite{tian2022scmelody}.

The consensus matrix, called $X$ (Fig. \ref{fig:multiple_results}G),
has optimal leaf ordering of its rows and columns. Such is produced by
the hierarchical clustering \textit{Ward} metric (also known as Ward's
minimum variance method), applied to the dissimilarity matrix $1-X$
(as seen in Fig. \ref{fig:multiple_results}F). This metric aims to
minimize the total within-cluster variance. The goal is to choose the
successive clustering steps to minimize the increase in the total
within-cluster variance. This metric is particularly effective for
creating clusters that are compact and have a roughly similar number
of elements \cite{ward1963hierarchical}, which was our main reason for
selecting it.

From this consensus, we can obtain various partitionings depending on
where the dendrogram is sliced.  When the dendrogram is sliced at a
value of 1.00 we obtain three major groups as observed in the thick
block structure in Fig.  \ref{fig:multiple_results}G.  Each block can
be further divided into smaller diagonal blocks if we reduced the
threshold. For instance, 7 clusters are observed for a cutoff at 0.55
represented by the smaller blocks in Fig. \ref{fig:multiple_results}G.
In the next sections, we will see that this partitioning is the optimal
solution obtained by the reconstruction algorithm with the least amount of
modifications to convert the baseline Varshney connectome into a
fibration symmetric solution (7 clusters for a cutoff at 0.55 in Table
\ref{tab:Percentages_and_pvalues}).

All these partitions are the input to the mixed integer linear
programming algorithm that we designed to reconstruct the connectome
with a minimal number of modifications, as explained next.

\begin{figure}[H]
  \centering
  \includegraphics[width=\linewidth]{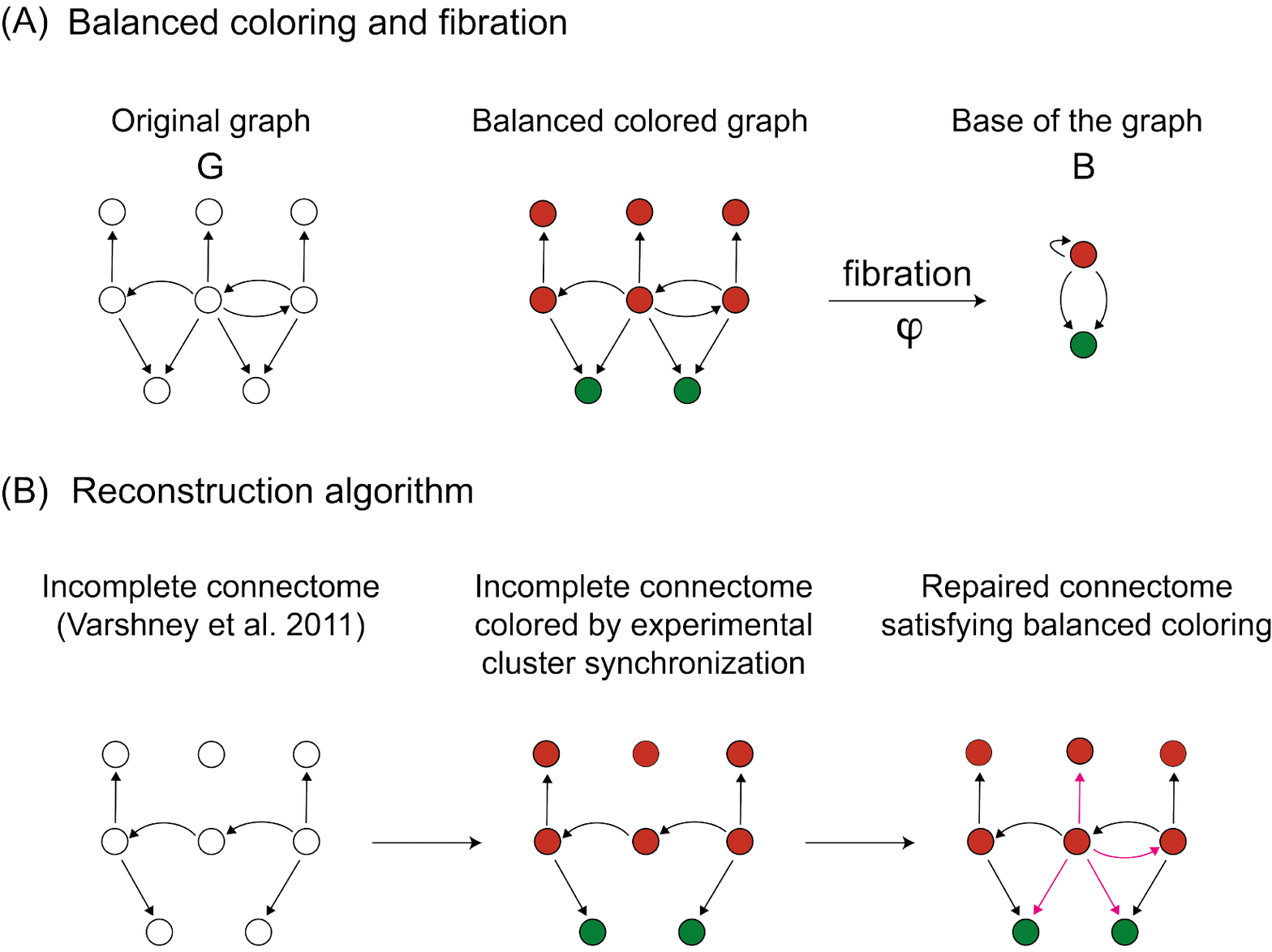}
  \caption{\textbf{Fibration formalism and reconstruction algorithm.}
    (A) Explanation of fibration symmetries and balanced
    colorings. The graph on the left has a (minimal) balanced coloring
    with 2 colors in the middle. For instance, each red node receives
    one red input and each green node receives two red inputs. The
    colored clusters represent clusters of synchrony when an
    admissible system of equations is superimposed on the graph. The
    graph $G$ can be collapsed by the symmetry fibration to the
    minimal base as shown in the right panel, where the edges satisfy
    the lifting property. The colored clusters (also called fibers)
    are the building blocks of the network. A refinement algorithm to
    obtain minimal balanced coloring is provided in
    \cite{morone2020fibration}. (B) SymRep reconstruction
    algorithm. We start with an incomplete connectome (left) to which
    colors are assigned (middle) according to the cluster
    synchronization obtained experimentally. A MILP then adds (or
    removes) a minimal number of edges (in red) to construct an ideal
    network (right) that satisfies the experimental balanced
    coloring.}
\label{fig:fibration}
\end{figure}

\subsubsection{Balanced coloring partitions, fibrations and
  cluster synchronization}
\label{sec:synchronization_via_fibers}

To develop the optimization algorithm to reconstruct the
connectome to satisfy the synchronization found in the previous
section, we need to introduce the concept of 'balanced coloring' and
the fibration of the graph.  Figure \ref{fig:fibration}A shows an
example of a graph with a balanced coloring.

Let $\mathcal{C} = {C_1, \cdot \cdot \cdot ,C_K}$ be a partition of
the nodes of a network $G=(V,E)$ with $V$ vertices and $E$ edges. We
identify each cluster $C_k$ with a different color and $K$ is the
total number of colors. A 'balanced coloring' is a coloring of the
graph such that each node with color $k$ in cluster $C_k$ is connected
(by edges of the same type) to the same number of nodes with color $j$
in cluster $C_j$, for $1\leq k, j \leq K$.  That is, nodes of the same
color receive the same colors from their neighbors
(Fig. \ref{fig:fibration}A, center)

Translated into the terminology of dynamical systems, a system of
differential equations for the state variables of each node $V_i(t)$
can be interpreted as a 'message passing' process of passing colored
messages through the edges of the graph. Since two nodes $i$ and $j$
with the same color received the same colors (messages) from their
neighbors, we can think, intuitively, that they will synchronize their
activity $V_i(t) = V_i(t)$, forming a synchronous cluster. This
intuition is made mathematically rigorous \cite{deville2015modular}
through the theory of fibrations
\cite{boldi2002fibrations,morone2020fibration}.

Traditionally, a way to formalize the balanced coloring partitions is
through the automorphisms of the graph forming its symmetry group
\cite{pecora2014cluster,sorrentino2016complete}. In this case, the
orbits of the automorphisms of the graph are the balanced
colors. However, not all balanced colorings are orbits. Many
biological networks contain no automorphisms, yet, they display a
non-trivial balanced coloring partition with many colors
\cite{morone2020fibration,leifer2020circuits}.  This is exemplified in
Fig. \ref{fig:fibration}A. The graph has no automorphisms (except for
the trivial identity), but has a balanced coloring that reduces the
graph to a base with just two nodes. This balanced coloring is only
captured by the fibration.

The graph fibration formalism introduced in \cite{boldi2002fibrations}
is a more general formalism that captures all balanced coloring
partitions of the graph. Graph fibrations are a particular case of
fibrations between categories introduced previously by Grothendieck
\cite{grothendieck1959technique} and are formal generalizations of
graph automorphisms.

A graph fibration (shown on the right of Fig. \ref{fig:fibration}A) is
a morphism of the graph that collapses every cluster of synchronized
balanced colors (called 'fibers') into a single representative node in
the base graph $B$ while conserving the 'lifting property' defined in
the Methods section \ref{sec:fibration}.  This transformation leaves
invariant the dynamics in the graph and captures the maximal
symmetries of the network. It is then called a symmetry fibration
\cite{morone2020fibration}. Here, we propose that fibers represent potential functional modules or building blocks in a network of neurons.



\subsection{Fibration symmetry driven repair algorithm} \label{sec:network_repair}

\begin{figure}[H]
  \centering
  \includegraphics[width=.89\linewidth]{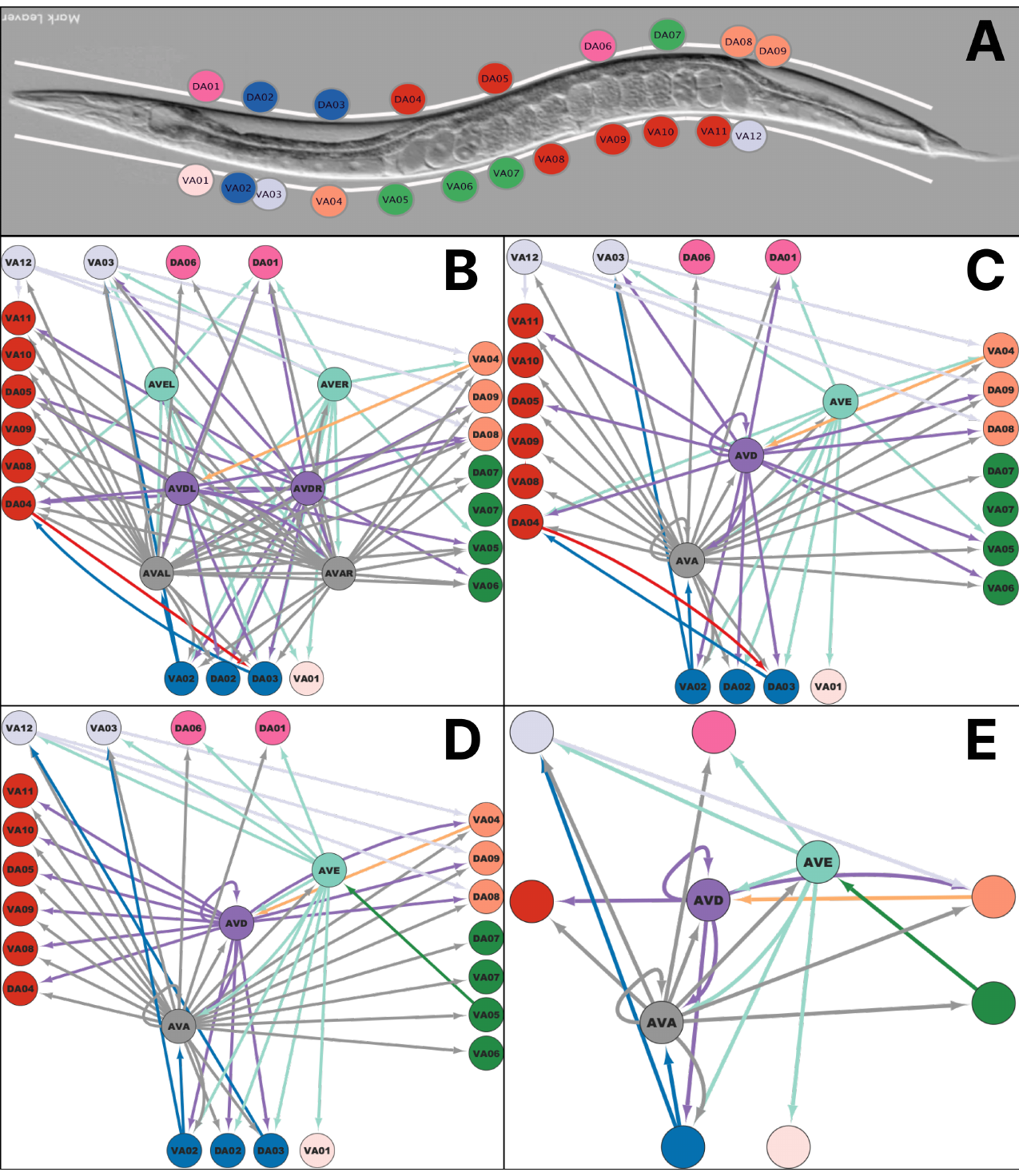}
\caption{\textbf{Reconstruction of the locomotion connectome.}  (A)
  Anterior-posterior distribution of motor neurons (DAn, VAn) involved
  in the backward locomotion in the \textit{C.elegans}
  \cite{hymanlab2012celegans} colored according to the most optimal
  consensus cluster synchronization method with cutoff 0.55 producing
  7 colors.  (B) Uncollapsed Varshney connectome from
  \cite{varshney2011structural}: Original Varshney chemical synaptic
  connectivity backward motor neurons and their three primary
  interneuron pairs AVAL/R, AVEL/R, and AVDL/R.  (C) Collapsed
  Varshney connectome: Left-right collapsing of the interneurons in
  the original network in (B). This is done by joining left and right
  interneuron pairs into one unit and applying OR logic on their
  outgoing edges, given by the pair of adjacency matrix values for
  those sharing the same target neuron for each left and right
  interneuron pair taken as the source neurons.  (D) Reconstructed
  network obtained by the MILP optimization algorithm, SymRep, applied
  to the network of (C) using the colors obtained from the
  synchronization analysis in Fig. \ref{fig:multiple_results}G. This
  network is perfectly balanced colored: all neurons belonging to a
  specific colored cluster receive the same amount of colored input
  from other neurons.  (E) The base graph representation of the
  network in (D).  }
\label{fig:consensus_solutions}
\end{figure}

The clusters of synchronous neurons found in section
\ref{sec:consensus_agreement} are used to repair the raw chemical
synapses connectome of the backward locomotion gait of
\textit{C.elegans} provided by Varshney {\it et al.}
\cite{varshney2011structural} (plotted in
Fig. \ref{fig:consensus_solutions}B). We develop an optimization mixed
linear integer program to modify this network by adding or removing
the least amount of edges within the limits of an objective function
\cite{wagler2014combinatorial} to match the synchronization obtained
experimentally. We call this algorithm the Symmetry-Driven Repair
Algorithm or SymRep for short.

The goal is to obtain a network with balanced coloring partitioning as
provided by one of the synchronization clusters from
Fig. \ref{fig:multiple_results} (details appear in the Methods section
\ref{sec:mip}). The most important feature of this algorithm in the
context of the present paper is the objective function used to
determine optimal solutions. Many combinations of adding and/or
removing edges can satisfy turning the connectome into a network with
color partitionings as provided by a functional clustering while
respecting the constraints provided by Eq. \eqref{eq:equality} and
Eq. \eqref{eq:unequality}. Of these many solutions, the one that
minimizes the objective function below provides an optimal solution
(Methods section \ref{sec:mip}):

    \begin{equation}\label{eq:cost_function_1}
    f_{\alpha,\beta}(r,a)=\alpha\sum_{i,j\in E} r_{ij} + \beta\sum_{i,j\in E^C} a_{ij}
    \end{equation}   
where $E$ is the set of edges present in the original connectome and
$E^C$ is the set of non-existing edges in the original connectome that
are permissible to be added. The terms $r_{ij}$ and $a_{ij}$ are
binary variables, where the first term, associated with $E$, takes on
a value of 1 if an edge has been removed. The second term, associated
with $E^C$, takes on a value of 1 if an edge has been added. If
necessary, one can control the SymRep algorithm to prohibit removing
(adding) connections between physically impossible neurons, which
studies show are always connected (disconnected). This can be done by
simply not including them in $E$ or $E^C$. The parameters $\alpha$ and
$\beta$ are penalty weight constants for each of these variables,
respectively. The relative weight between these parameters determines
if the SymRep prefers to repair a network through the addition of
edges rather than removal. In our study, we explore the solutions
found by varying the relative differences between $\alpha$ and
$\beta$. We keep the value of $\alpha$ at one and increment the value
of $\beta$ in steps of one, starting from one and culminating at 10.

We work with the raw chemical synapses connectome of
\textit{C.elegans} for the backward locomotion gait provided in
\cite{varshney2011structural}. This is composed of the connectivity
between motorneurons (VAs and DAs) and 3 main interneuron pairs
(AVAL/R, AVEL/R, AVDL/R). We specifically use two versions of this
connectome.  One is that reported by Varshney {\it et al.}
\cite{varshney2011structural} shown in
Fig. \ref{fig:consensus_solutions}B.
We call it the Uncollapsed Varshney connectome. The second, shown in
Fig. \ref{fig:consensus_solutions}C, is a collapsed version of this
network in which we assume that motorneurons can not distinguish
between signals received from the left or right version of an
interneuron pair (i.e., AVAL or AVAR). This is done by substituting an
interneuron pair for one node (i.e., AVAL and AVAR are collapsed into
AVA) and substituting the two edges a neuron may receive from an
interneuron pair for one edge with a weight equal to 1 if both edges
are present, one if only one edge is present, and 0 if not
present. This version is called the Collapsed Varshney connectome. As
a final note, we assume each interneuron pair belongs to its unique
synchronization cluster.

\subsection{Synchronization-driven reconstruction of the connectome} \label{sec:repair_solutions}

With the 39 unique clusterings found in the previous section, the 19
different combinations of weight penalties and the two different raw
connectomes (original and collapsed versions) used, we run SymRep to
produce 1,482 solutions. All of these solutions satisfy the condition
of balanced coloring. Of the 741 solutions for the collapsed Varshney
connectome, 730 (98.52\%) satisfy the stricter condition of minimal
balanced coloring, and 737 (99.46\%) fulfill this condition for the
not collapsed version. Of these, we present the most optimal solution
for each type of cluster or community detection algorithm in
combination with the functional measurement group ($LoS$ and
correlations).

We decide on an optimal solution based on the number of modifications
it enacts on the raw connectome with the condition that this number is
the lowest, and at least below ~50\% of natural variability found from
animal to animal \cite{witvliet2021connectomes, hall1991posterior} to accept the
solution. We pick those that modify the connectome by the least amount
of addition and removal of edges. We consider two cases to measure the
number of modifications. The first one is where adding or removing has
a cost of 1, where the final sum of modifications is divided by the
total number of edges in the raw connectome. The second case is in
which the severity of  
removing an edge is tied to the weight of the
edge in the collapsed connectome before binarization while adding an edge only has
a cost of one.  In the latter, the total sum of modifications
(additions + removals) is divided by the total number of edges in the
raw connectome. Metadata about the solutions can be observed in Table
\ref{tab:Percentages_and_pvalues} with significant p-values for each method.

\subsection{Statistical significance}\label{sec:statistical_significance}

\begin{figure}[h]
   \centering
  \includegraphics[width=\linewidth]{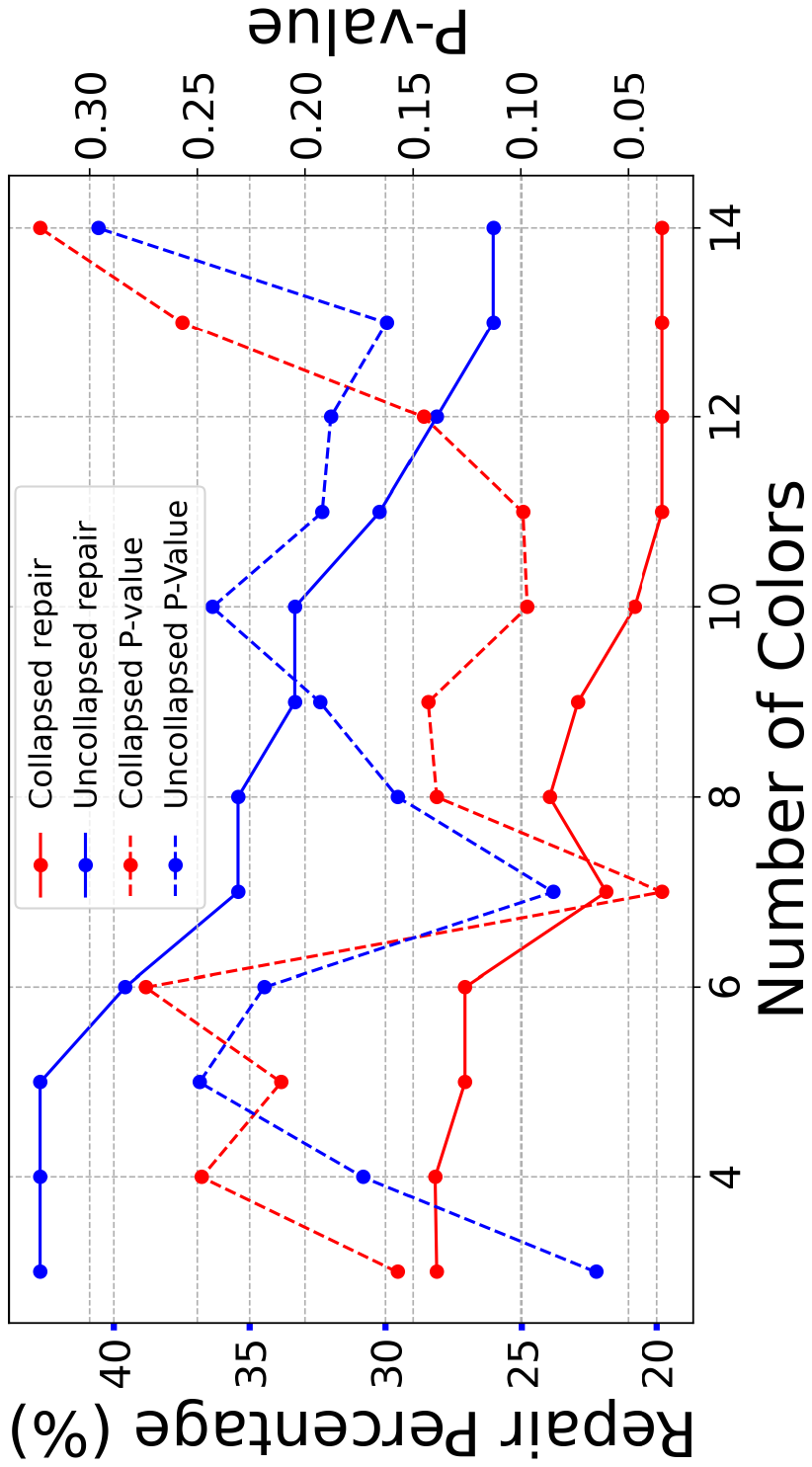}
   \captionsetup{font=small}
   \caption{{\bf Statistical analysis.}  The plot visualizes the
     percentage of edge modifications required to achieve fibration
     symmetry in networks as a function of node consensus partitioning into
     \(N\) clusters, represented along the x-axis. Some of these can
     be seen in Figs. \ref{fig:multiple_results}H-I. The graph
     compares repairs done on the 'Collapsed' and 'Uncollapsed'
     backward gait network of the \textit{C.elegans} with the
     percentages needed for different numbers of colors
     (clusters). Additionally, we show the statistical significance
     (P-values) of these solutions against 1,000 permutations with
     similar cluster distributions but shuffled node labels. The dual
     y-axis format highlights the repair percentages (continuous
     lines) and their P-values (dashed lines), providing a
     comprehensive view of the network modifications required to
     achieve fiber symmetry in this network. The lowest p-value (and
     below the standard 0.05 p-value cutoff for acceptance) with
     repairs as low as 22\% are found for 7 clusters at consensus
     clustering cutoff of 0.55.}
\label{fig:Percentages_and_pvalues}
\end{figure}

To test the significance of these functional partitionings and
eventually choose the most significant partition, we proceed to use a
permutation test \cite{good2013permutation}. Specifically, for each of
the partitionings considered in Figure
\ref{fig:Percentages_and_pvalues} we randomly permute the labels
(names) of the neurons, keeping the number of clusters and number of
nodes distributed among these the same. This leads to a shuffling of
synchronous groups keeping their size (amount of neurons in each) the
same. We produce slightly over 1,000 permuted versions of each of
these partitionings and inspect how many lead to networks with these
partitionings as fibers with equal or less number of edge
modifications. This quantity is then divided by 1,000 or by the number
of networks that lead to a minimally balanced coloring solution with
an equal number of fibers as in the partitionings ($\sim$1,000).


\begin{table}
\begin{tabular}{|c|c|c|c|c|c|c|c|} 
\hline
Algorithm & Measure & Value & Modified & $\alpha$ & $\beta$ & Fibers & P-value \\ 
\hline
\multicolumn{8}{|c|}{\textbf{Individual partitionings applied on Uncollapsed Varshney connectome}} \\ 
\hline
Louvain & Corr. & Cov. & 42.71\% & 1 & 1 & 2 & 0.023 \\ \hline
Louvain & $LoS$ & 0.06 & 38.54\% & 1 & 1 & 3 & 0.007 \\ \hline
Clique sync & Corr. & Cov. & 40.62\% & 3 & 1 & 2 & 0.003 \\ \hline
Clique sync & $LoS$ & 0.16 & 38.54\% & 1 & 1 & 3 & 0.008 \\ \hline
\multicolumn{8}{|c|}{\textbf{Individual partitionings applied on Collapsed Varshney connectome}} \\ 
\hline
Louvain & Corr. & Cov. & 26.04\% & 1 & 2 & 2 & 0.019 \\ \hline
Louvain & $LoS$ & 0.06 & 26.04\% & 1 & 2 & 3 & 0.033 \\ \hline
Clique sync & Corr. & Cov. & 25.00\% & 1 & 1 & 2 & 0.003 \\ \hline
Clique sync & $LoS$ & 0.17 & 23.96\% & 1 & 1 & 3 & 0.007 \\ \hline
\multicolumn{8}{|c|}{\textbf{Consensus partitionings applied on Collapsed Varshney connectome}} \\ 
\hline
Consensus & Cutoff @ & 1.73 & 28.12\% & 1 & 2 & 3 & 0.157 \\ \hline
Consensus & Cutoff @ & 0.85 & 28.12\% & 1 & 2 & 4 & 0.248 \\ \hline
Consensus & Cutoff @ & 0.77 & 27.08\% & 1 & 2 & 5 & 0.211 \\ \hline
Consensus & Cutoff @ & 0.58 & 27.08\% & 2 & 1 & 6 & 0.274 \\ \hline
Consensus & Cutoff @ & 0.55 & 21.88\% & 1 & 1 & 7 & 0.033 \\ \hline
Consensus & Cutoff @ & 0.50 & 23.96\% & 1 & 1 & 8 & 0.139 \\ \hline
Consensus & Cutoff @ & 0.40 & 22.92\% & 1 & 1 & 9 & 0.143 \\ \hline
Consensus & Cutoff @ & 0.34 & 20.83\% & 1 & 1 & 10 & 0.099 \\ \hline
\end{tabular}
\caption{{\bf Solution metrics.}  The first two blocks in this table
  correspond to the best solutions found for the intersection between
  the community detection algorithm and the metric used to determine
  synchronization (Correlations vs $LoS$). The third block, located at
  the bottom corresponds to the first number of partitionings for the
  consensus method as visualized in Fig. \ref{fig:multiple_results}F
  with an emphasis on showing the values before and after the most
  optimal partitioning found across all tested partitionings (cutoff
  at 0.55). The p-values indicate how often the partitioning performs
  equally or better than versions of the same partitioning with the
  labels of its nodes randomly shuffled. The values of $\alpha$ and
  $\beta$ indicate the penalty values for which the algorithm found
  the optimal solution. The fiber column indicates the number of
  clusters for the given partitioning among motor neurons. These
  numbers do not include the three rigid clusters reserved for the
  inter-neurons.}
\label{tab:Percentages_and_pvalues}
\end{table}

\subsection{The optimal solution}
\label{sec:optimal_clusters}

The consensus matrix allows the detection of fibers at different
scales (number of clusters) through the information contained in its
hierarchical clustering, as seen in
Fig. \ref{fig:multiple_results}F. These fibers are used for repairing
the Varshney chemical synaptic backward gait produces some outstanding
results among the collapsed and non-collapsed connectome. We highlight
the results for the collapsed connectome at the bottom portion of
Figure \ref{fig:Percentages_and_pvalues}.

\smallskip
When it comes to the statistical significance of clusters obtained
through a system of ever-increasing node partitionings to repair a
network, as in our case, there is some expected behavior. For
instance, the trivial partitioning of $N$ neurons into $N$
non-overlapping clusters will give a p-value of 1 as all permutations
of a trivial partitioning will produce the same partitioning,
resulting in all of them modifying the network by the same amount. The
same can be said for a trivial partitioning with 1 cluster as all
permutations result in the same partitioning. Therefore, as the number
of clusters change between 1 to $N$, the p-value of statistical
significance is granted to produce a (global) minimum p-value between
1 and $N$ clusters \cite{good2013permutation}.  For a low-noise system
in which node dynamics are faithful to the network structure with
$1\le M\le N$ fiber partitionings, it is suggested that as the number
of non-overlapping clusters is reduced from $N$, the p-value is
expected to improve (be below 1), with the same expected p-value as
the number of clusters is increased above one where the minimum
p-value should be located at $M$ clusters. Deviations from this
depends on how far removed the network under repair is from the
original fiber symmetric version producer of the node mentioned above
dynamics. These predictions in the behavior of p-value are observed in
the repairing of the backward locomotion network in
Fig. \ref{fig:Percentages_and_pvalues}. As the number of clusters
approaches the trivial $N$-partitioning the amount of repairs needed
decreases because the network we are trying to repair is indeed
originally trivially colored.

Applying all these considerations, we find that the most optimal
partition and reconstructed connectome is obtained for 7 clusters at
consensus clustering cutoff of 0.55 repaired from the collapsed
Varshney connectome.  As shown in
Fig. \ref{fig:Percentages_and_pvalues} and Table
\ref{tab:Percentages_and_pvalues} this solution has the lowest
p-value=0.033, being the only one of the consensus solutions below the
standard 0.05 p-value cutoff for statistical acceptance. Furthermore,
the repair percentage is 21.88\%, below the acceptance limit of 25\%
variability from animal to animal. Therefore, we choose this
connectome, shown in Fig. \ref{fig:consensus_solutions}A, as the one
representing the data in the closest possible way.

\section{Discussion}
\label{discussion}


Understanding how the structure of the connectome influences the
function of the network is a long-standing problem in system
neuroscience \cite{sporns2013structure}. However, addressing this
question is challenging due to low sample number in available connectome datasets and potential 
variability across individuals.  We have addressed the problem of incomplete/missing data in
biological networks by first developing a consensus method to obtain
average clusters of neuron synchrony across worms from whole-body
calcium recording of neural activity. This information drives a
reconstruction optimization algorithm of the connectome consistent with the observed synchronization.  When the modifications to
the connectome are below the experimentally found variation from
animal to animal and offer a
minimal p-value significance, the obtained connectome models can be thought
of as idealized networks consistent with the experimentally measured dynamics of the neuronal circuits. Such a network bridges the gap between structure
and synchronization and can be used to further assess the
functionality of the connectome via perturbations to its structure.

Our research has substantiated the pivotal role of fibration
symmetries in the connectome structure in orchestrating neuronal
synchronization in \textit{C. elegans}. The refined understanding that
structural connectomics underpins significant aspects of neural
functionality furthers our grasp on the physical bases of neural
synchronization. This synchronization is crucial, as it forms the
foundation for coordinated motor outputs and behavioral responses in
organisms. The use of advanced calcium imaging and graph theory to
correlate these structural motifs with dynamic patterns of neural
activity offers a compelling model for predicting neuronal behavior
based on underlying anatomical data.

The methodological innovations in our study, particularly the use of
integer linear programming (SymRep) to adjust connectomics data based
on functional imaging, demonstrate a novel computational approach to
neuroscientific research. This technique refines existing neural
network models and provides a quantifiable method for aligning
theoretical predictions with empirical observations.

Previous research in {\it C. elegans} has suggested local oscillators at the motor
neuron level for rhythm generation in the VNC for reverse
locomotion \cite{gao2018excitatory}; these patterns have been observed when motor neurons were experimentally deprived of their descending interneuron inputs. Our experimental conditions, Ca++ imaging in immobilized worms with interneurons AVA and AVE generating descending motor commands, did not reveal any oscillations or activation patterns consistent with a rhythmic backward gait pattern i.e., alternations between VA- and DA-motorneurons, or a traveling wave from the posterior to anterior position. Instead, we observed that A-motor neurons faithfully mirror the descending motor command activity in sync. It was shown that {\it C. elegans} can adapt their gait flexibly to the physical properties of the environment via proprioception, as shown for forward locomotion \cite{wen2012proprioceptive}, and that A-motorneurons have indeed proprioceptive properties themselves  \cite{zhan2023locomotion}. We suggest that proprioception might be an essential driver to unleash the oscillatory or rhythmic properties observed in ref.  \cite{gao2018excitatory}. In our study, the lack of proprioceptive inputs and absence of complex internal motorneuron dynamics was an advantage, since it enabled us to study the activity of all A-motorneurons solely in response to their synaptic partners, providing a filter on activity that should be mostly explainable by the connectome architecture alone.   

Connectomes between adult worms are remarkably variable, in particular for connections between neurons with few synaptic contacts \cite{witvliet2021connectomes} as well as in the posterior ventral cord (VNC) \cite{hall1991posterior}. Extrapolating from these data, connections between descending interneurons and motor neurons in the entire VNC might vary by up to 50\% i.e., about half of the connections found might not be reproducible across individuals. Our connectome repair algorithm provides a family of statistically significant solutions that require connection additions/removals well below this range (Table 1, Fig. 6). All of these solutions provide circuit models with fiber symmetrical properties consistent with synchronization patterns found in A-motorneurons. It is unlikely, that individual animals indeed perfectly match these idealized symmetric solutions, thus neuronal networks must be endowed with additional properties that ensure stable synchronous dynamics. However, here we propose that fibration symmetries provide constraints, from which individuals should not deviate too much to ensure the robustness of synchronous dynamics. Here we show that symetrization procedures can, however, be useful, for identifying potentially functional circuit modules in the resulting base graph. 

Our analyses led to a base graph of the backward locomotion circuit that suggests functional modules characteristic of differential innervation by the descending interneurons AVA, AVE, and AVD (Fig. \ref{fig:consensus_solutions}E), which intriguingly appear as a crude topographic map (Fig. \ref{fig:consensus_solutions}A). Our analyses thereby suggest that these modules could operate as functional units in deferentially transmitting the motor commands to different body parts, perhaps in a manner supporting effective backward locomotion. AVA, AVE, and AVD differ to some degree in their connectivity to other sensory circuits \cite{white1986structure}; we thus speculate that the modules might enable differential control over how to execute the reversal motor command. Note, that AVD neurons were not active in our recordings. Future studies that activate also AVD, and selective inhibition or interference with AVA, AVE, and AVD will test the relevance of our findings for the control of A-motorneuron activity and backward crawling behavior.   

The present study focused on a smaller neuronal circuit where some of the assumptions that our theory includes might be reasonable simplifications. Anatomically and molecularly, left and right members of interneuron classes AVA, AVE and AVD are nearly indistinguishable \cite{taylor2021molecular}, justifying bilaterally collapsing the connectome (Fig. \ref{fig:Percentages_and_pvalues} and Table
\ref{tab:Percentages_and_pvalues}). Moreover, most chemical synapses in the circuit are excitatory and cholinergic \cite{pereira2015cellular}, and VA and DA motorneurons have similar molecular footprints \cite{taylor2021molecular}, justifying the simplification of node and link equivalency. Future work on larger circuits in the worm and other model organisms should develop our theoretical approach further to realistically allow for more diversity in neuronal and synaptic properties.  


Indeed, the implications of our findings are not restricted to
\textit{C.elegans}. The principles of neuronal synchronization,
facilitated by connectomic structures observed in this study, likely
have analogs across bilaterian species, including humans
\cite{gili2024fibration}. Investigating these principles in more
complex nervous systems could reveal new insights into how brain-wide
synchronization patterns contribute to complex behaviors and cognitive
functions. Furthermore, exploring the impact of structural variations
on the synchronization in disease models could open new therapeutic
avenues, particularly for neurological disorders where dysregulation
of synchronized activity is evident.

\clearpage
\section{Methods} \label{Methods}

In this section, we first explain the experimental setup and the
methods to extract the functional synchronization clusters of neurons
obtained using various combinations of functional measurements and
different methods of extracting clusters and communities. Following
this, we introduce the graph fibration formalism to describe balanced
coloring partitions of graphs and cluster synchronization. We conclude
with the repair algorithm for optimal fibration symmetric network
solutions (SymRep) applied to the synchronization partitionings.


\subsection{Neuronal calcium imaging} \label{sec:apparatus_and_setups}

Whole-nervous system Ca2+ imaging experiments were performed on
transgenic young adult \textit{C.elegans} hermaphrodites (age was
determined by the number of a maximum of 5 eggs) expressing genetically-encoded
calcium indicator NLS-GCaMP6f in a pan-neuronal fashion and localized
to the cell nuclei \cite{kaplan2020nested, Uzel2022neuronhubs}
together with NeuroPal cell class identification labels
\cite{yemini2021neuropal} (ZIM2001: \textit{otIs669; MzmIs52 ; lite-1 (ce314)})

\smallskip
{\bf Mounting and microfluidic setup:} Animals were imaged in
two-layer PDMS microfluidic devices to control the oxygen environment
and with curved channels \cite{Uzel2022neuronhubs} to immobilize and
laterally align animals, enabling reliable positioning of worms across
recordings and fitting them into the field of view of the imaging
system. Such a layout allowed us to cover the head ganglia, ventral
cord, and tail ganglia of the animals.  The worm channel of the
microfluidic device was connected to a syringe that contains NGM
buffer with 1 mM tetramisole to paralyze worms. All components were
connected using Tygon tubing (0.02 in ID, 0.06 in OD; Norton) using
23G Luer-stub adapters (Intramedic). Constant gas flow of 21\% O2 and
79\% N2 (50ml/min) was delivered using a gas mixer connected to mass
flow controllers (V\"ogtling Instruments) using lab custom scripts in Micromanager. Adult worms were picked on food-free NGM agar plates in a
drop of NGM with 1mM tetramisole and aspirated into the worm channel.
Animals were first habituated for 10min and afterward imaged at 21\%
O2 for 10 min.

\smallskip
{\bf Microscope setup:} High-resolution data of neuronal activity in
the head and tail ganglia were acquired with an inverted spinning disk
confocal microscope (Zeiss Axio Observer.Z1 with attached Yokogawa
CSU-X1) using a sCMOS camera (pco.edge 4.2 with Camera Link HS connection) and a 40x 1.2 LD LCI Plan- Apochromat
water-immersion objective (Zeiss). Moreover, we used 0.5x demagnification relay optics at the port of the spinning disk unit to increase intensity/pixel and the total area that can be imaged on the camera sensor both by the factor of 2, allowing imaging of a larger field of view encompassing the full worm nervous system. We use a custom-made GUI in
Micromanager to control the different elements of the
microscope. Exposure time was 20 ms, with 2$\mu$m steps between Z-planes
operated by a Piezo stage (P-736 PInano, Physik Instrumente GmbH); the
total plane number of z-planes varied between 16-20, leading to a
volume acquisition rate of up to 3 Hz.

\smallskip
{\bf Neural trace extraction:} As described in detail in our previous
work in Kato {\it et al.} \cite{kato2015global}, neuronal activity
traces were obtained by tracking the intensity maxima in each volume
over time and calculating the single-cell fluorescence intensities. F0
was calculated for every neuron as the mean fluorescence intensity
across each trial. After background subtraction, DF/ F0 was calculated
for each neuron, following bleach correction by linear detrending and
exponential fitting. See reference \cite{kato2015global} for details.

\smallskip
{\bf Neuronal identification:} Identification of each neuron was done
based on a neuron dictionary termed NeuroPAL \cite{yemini2021neuropal}
(Fig. \ref{fig:neuroPAL}A).  In each recording, we aim to detect at least 25 crucial neurons for our analysis: two reversal interneurons AVA and AVE, and all the reversal motor neurons DA01-DA09 and VA01-VA12.  Some neurons in the head and tail
tip were lost in individual recordings, depending on animal size, when it exceeded the imaging
region. Other reasons for missing neurons might be a failure in
segmentation and tracking due to low signal/noise ratios based on low
expression levels and/or low calcium levels. Neuronal traces were
curated after each recording, and obviously erroneous traces were
removed.
Average correlation matrices were generated by calculating the mean pairwise correlations between the activity time series of identified active neurons (up to n=8, depending on the number of pairwise observations, but at least n=3). The identified active neuron numbers are as follows: 72 for recording 1, 76 for recording 2, 98 for recording 3, 79 for recording 4 and 5, 69 for recording 6, 63 for recording 7, and 62 for recording 8

{\bf NeuroPAL labeling:} The identities of neurons were determined via
NeuroPAL using the following procedure. We obtained image stacks from
each recorded animal after GCaMP6f imaging.  For each plane, we
acquired spectrally isolated images sequentially of CyOFP1,
mNeptune2.5, and mTagBFP2. We excited CyOFP1 using the 488nm laser at
5\% intensity with a 585/40 bandpass emission filter. Afterward,
mNeptune2.5 and TagRFP-T were recorded using a 561nm laser at 20\% intensity (percentage of max) with a 655LP filter and a 570LP filter, respectively. mTagBFP2 was isolated using a 405nm laser at 30\% intensities with a 447/60 bandpass filter. These imaging
conditions allow for the color-specific labeling of
\textit{C. elegans} neurons. Neuronal identities were manually annotated
according to the NeuroPal guidelines. \cite{yemini2021neuropal}.

\subsection{The zoo of synchronization measures}  \label{sec:synchronization_measures}

Determining the synchronization between two or more simultaneously
recorded signals is a task with multiple approaches depending on the
characteristics that one is interested in measuring and determining
their level of similarity. Naturally, all this leads to a zoo of
methods that try to capture the synchronization between signals
through different features \cite{kreuz2013synchronization}. These
methods tend to be broken into four main camps of measurement, which
are the intersection of the time domain vs the frequency domain and
those that consider directional dependencies and that do not; some of
these examples can be observed in Table
\ref{table:synchronization_measures}.

\begin{table}
\centering
\begin{tabularx}{\textwidth}{|c|X|X|} 
\hline
 & \textbf{Time Domain} & \textbf{Frequency Domain} \\ \hline
\textbf{With Directional Dependency} & Cross-correlation, Granger Causality, Transfer Entropy & Phase Locking Value, Directed Transfer Function, Partial Directed Coherence \\ \hline
\textbf{No Directional Dependency} & Pearson Correlation, Spearman's Rank Correlation, Mutual Information & Coherence, Spectral Correlation, Phase Coherence \\ \hline
\end{tabularx}
\caption{Synchronization Measures in Time and Frequency Domains}
\label{table:synchronization_measures}
\end{table}

Depending on the data and the aspects of the signals one is interested
in, one may use a few examples in Table
\ref{table:synchronization_measures} and even other methods not
mentioned in the table. Selecting the appropriate metrics can
sometimes be a difficult problem, and one must compare these against
some reference, producing competition between the selected
metrics. Below, we indicate which metrics we have decided to
use. Further, down in the paper, it is shown that it is best not to
put these metrics in competition one against the other to find the
best metric that captures synchronization. Instead, a better approach
is to use the information that these provide in a consensus manner to
decipher the synchronization in our system \cite{tian2022scmelody}.

\subsection{Clique synchronization and Louvain method} 
\label{sec:synchronization_used}

From section \ref{sec:functional_clusters}, the Clique Synchronization
method developed in \cite{gili2024fibration} accepts a clique if all
nodes composing a clique satisfy the conditions outlined in section
\ref{sec:functional_clusters}. We define a cluster of $N$ neurons to
be synchronous as a fully connected clique consisting of $N$ nodes
that meet the conditions:
\begin{equation}
\sum_{i<j}^{1,N} \sigma(x_i (t),x_j(t)) \geq \frac{N(N-1)}{2} \sigma(x_k (t),x_{k'}(t)) \quad \forall k= 1, \ldots, N \text{ and } k' \in \mathcal{M}_k, 
\end{equation}
where $\mathcal{M}_k$ denotes the set of nearest neighbors of node $k$
(for $k=1, \ldots, N$) that are not part of the clique in
question. Here, $\sigma(x_i (t),x_j(t))$ represents the value for
synchronization, measured by the $LoS$ metric or any correlation
metric used in this paper.

\smallskip

\smallskip
The Louvain method \cite{blondel2008fast} is a popular algorithm for
module and community detection. At every execution, it assigns every
neuron to its community, as per the method used here, it proceeds to
randomly group neurons into bigger clusters, only accepting mergers if
the modularity value given by Eq. \eqref{eq:modularity} increases and
halting once this equation can no longer increase in value
\cite{blondel2008fast}. Due to the stochastic nature of this process,
each execution may lead to a slightly different result. Due to this,
the Louvain method is executed 1,000 times for a given network
retaining the partitioning with the highest measured modularity. The
modularity measure is defined as:
    \begin{equation}\label{eq:modularity}
    Q=\frac{1}{2m}\sum_{i,j}[A_{ij}-\frac{k_{i}k_{j}}{2m}]\delta(c_{i},c_{j}),
    \end{equation}
where $A_{ij}$ is the adjacency matrix of the network, $m$ is the
number of edges in the network, $k_i$ is the number of in-degree
edges attached to node $i$ and $c_{i}$ are numeric labeled given to
each synchronization cluster \cite{blondel2008fast}.

\subsection{Fibration symmetry of graphs and cluster synchronization}
  \label{sec:fibration}

In the study of complex networks, understanding the interplay between
structure and dynamics play a crucial role. Specifically, the concept
of synchronization, where nodes in a network adjust their behavior in
accordance with each other is fundamental in various disciplines,
including physics, biology, and engineering. This section delves into
the intricate relationship between graph topology, characterized by
fibration symmetry, and the emergence of synchronization through the
lens of admissible ordinary differential equations (ODEs)
\cite{stewart2003symmetry,golubitsky2006nonlinear,morone2020fibration}.

A graph $G=(V_G,E_G)$ consists of a set of vertices $V_G$ and a set of
edges $E_G$ where each edge connects a pair of nodes $u$ to $v$
represented as $(u,v) \equiv e^{u \rightarrow v}_G$. The set of edges
from node $u$ to node $v$ is written as $E_{G}(u,v)$, that is, the set
of edges $e \in E_G$ that have as a source node $u$ brought by the
function $s(e)=u$ and have as a target node $v$ brought by the
function $t(e)=v$. The set of edges in graph $G$ that have as a target
the node $v$ is denoted by $E_{G}(-,v)$ \cite{diestel2017graph}.

\smallskip
A graph $G$ can be mapped into another graph $W$ through the structure
preserving process of a morphism $\Psi:G \longrightarrow W$ using a
pair of functions $\Psi_V:V_{G} \longrightarrow V_{W}$ and
$\Psi_E:E_{G} \longrightarrow E_{W}$. These functions map vertices to
vertices and edges to edges, respectively, from graph $G$ to $W$,
while being commutative with the source and target mapping
functions. In mathematical terms, these functions must obey
\cite{boldi2002fibrations}:
\begin{equation}
s_{W} \circ \Psi_{E} =
\Psi_{V} \circ s_{G} ,
\end{equation}
and
\begin{equation}
  t_{W} \circ \Psi_{E} = \Psi_{V} \circ
  t_{G} .
\end{equation}

If the mapping functions $\Psi_V$ and $\Psi_E$ are surjective, that is
to say, they map multiple elements (vertices or edges) in their domain
(graph $G$) to one representative element in their range (graph $W$),
then morphism $\Psi$ is called an epimorphism.

\smallskip

A (surjective) graph fibration is a particular type of epimorphic
morphism that maps elements that belong to the same category into one
representative element while preserving the lifting property which
guarantee dynamical invariance.  The original definition of fibration
was defined between categories by Grothendieck and others
\cite{grothendieck1959technique}. Boldi and Vigna
\cite{boldi2002fibrations} worked out the definition of fibrations
between graphs, which is the main theoretical framework of the present
work.  Morone {\it et al.}  \cite{morone2020fibration} then showed the
application of fibration to biological networks to understand their
building blocks \cite{leifer2020circuits}, and synchronization
\cite{leifer2021predicting,avila2024fibration}.

A \textbf{graph fibration} is a morphism from
graph $G$ to the base  $B$:
\begin{equation}
  \varphi: G \longrightarrow B
\end{equation}
that satisfy the lifting property \cite{boldi2002fibrations}.  This
means that for every edge $e = (u,v) \in E_G$, there is an edge $e' =
(\varphi(u), \varphi(v)) \in E_B$, and for every vertex $v \in V_G$ there
exists a unique edge $e^{v} \in E_{G}(-,v)$ satisfying $(w,\varphi(v))
\in E_B$.

The importance of the lifting property is that the dynamics between
the graphs $G$ and $B$ is preserved. This means that if we add a set
of admissible ODEs to the graph with dynamical variables $x_i(t)$,
then the dynamical evolution of the system in graph $G$ is the same as
the dynamical evolution of the graph $B$.
\begin{equation}
  \mbox{\rm Dyn}(G) =   \mbox{\rm Dyn}(B) .
\end{equation}
This property defines the clusters of synchrony as follows.

\smallskip
When a function $\varphi: G \rightarrow B$ acts as a fibration, we
refer to $G$ as the total graph and $B$ as the base graph of this
mapping function. In this context, $G$ is said to be fibred over
$B$. The set of vertices in $G$ that $\varphi$ sends to a specific
vertex $x$ in $B$ is called the fiber over $x$, denoted by
$\varphi^{-1}(x)$. Fibers are also balanced color clusters.

Vertices belonging to the same fiber have isomorphic input trees as
defined in \cite{morone2020fibration}. The input tree for a node $v$,
denoted $T(v)$, is a rooted tree centered at node $v$ that captures
all the paths in the graph leading to $v$.  The first layer of the
tree is the node's in-neighborhood, called its input set.  Each
subsequent layer is then iteratively defined as the input set's input
set. A visual example of an input tree can be seen in
Fig. \ref{fig:pipeline}I, where all blue nodes in the total space
graph have the same input tree structure.

In terms of dynamics, we can think of a message passing process where
the information travels through the input tree and arrives at the
rooted node $v$. If another node $u$ has an isomorphic input tree with
$v$, meaning that $T(v) \sim T(u)$, then these two nodes receive the
same messages through the network, although from different
pathways. Therefore $u$ and $v$ synchronize their dynamics $x_v(t) =
x_u(t)$. This statement has been put in rigorous mathematical terms by
DeVille and Lerman \cite{deville2015modular}.

A fiber is called trivial if it contains exactly one vertex, that is
if $|\varphi^{-1}(x)| = 1$ and called nontrivial if
$|\varphi^{-1}(x)|>1$. Throughout this paper, the function $|\cdot|$
denotes the number of elements in the mathematical structure it
encloses unless specified otherwise.

\smallskip

Vertices belonging to the same fiber, say $u$ and $v$, have an
equivalence relation $\simeq$ called an in-isomorphism, which can be
seen as a particular case of a discrete homeomorphism
\cite{boldi2002fibrations}, where a discrete bijective function
(one-to-one association of elements in two domains) $\psi: G(-,u)
\rightarrow G(-,v)$ is applied to the discrete topological space of
graphs.

Additionally, the \textbf{in-} in in-isomorphism is the added
condition for the bijective function to obey $s(e) \simeq s(\psi(e))$
for all $e \in G(-,u)$. In short, in-isomorphism holds the notion that
the vertices in a graph $G$ can be converted into one another while
conversing the adjacency connectivity from the same (or equivalent)
source vertices.

\smallskip
A fiber in a graph $G$ is associated with the cluster containing the
vertices of the fiber $u, v, w,...\in C_i$ where $n_i$ represents the
number of vertices in cluster $C_i$ with its vertices noted as
$\cup_{l=1}^{n_i} v^{i}_{l} = C_i$. The union of these clusters
contains the entirety of vertices $C_1 \cup C_2, \cup...C_K\in V_G$
where $K$ is the number of fibers in graph $G$. The overlap of
different clusters is always empty $C_i\cap C_j=\emptyset$ for $i\neq
j \; \forall i,j\in K_{G}$ where $K_{G}=\{x\in \mathbb{Z} | 1 \leq x
\leq K \}$ such that $\sum_{c=1}^{K} n_c = |V_G|$.

These fibers are the equitable partition or colored-balanced
partitioning of the graph as the number of in-degree edges of the
in-isomorphic vertices in a cluster $C_i$ receives from another
cluster $C_j$ only depends on the choice of the clusters.  Thus, both
fibers and balanced colorings describe, in two different ways, the
same synchronous dynamics of clusters of nodes.

We can represent this as:

\begin{equation}
C_{i} \leftarrowtail C_{j}\equiv\{|\cup_{l=1}^{n_j} E_{G}(v_{l}^{j},
v^{i})|=|\cup_{l=1}^{n_j} E_{G}(v_{l}^{j}, v^{k})|\}\; \,\,\,\,\,
\forall v^{i},v^{k}\in C_i
\label{eq:equality}
\end{equation}
where the number of edges $|\cdot|$ every vertex in $C_i$ receives
from a cluster $C_j$ must the same \cite{kudose2009equitable}.

\smallskip

The particular type of fibration we work with in this paper is the
\textit{minimal} fibration, which leads to a minimal equitable
(colored balanced) graph partitioning. The word minimal enforces the
value $K$ representing the number of colored balanced partitionings of
a graph to be the lowest it can be. This leads to the situation in
which no two clusters receive the same in-degree edges relative to a
third cluster, which leads to the constraint
\begin{equation}
  \sum_{k=1}^{K} |(C_{i} \leftarrowtail C_{k})-(C_{j}\leftarrowtail
  C_{k})|>0\; \,\,\,\,\, \forall i,j\in K_{G}.
\label{eq:unequality}
\end{equation}

The surjective minimal graph fibration is called a symmetry fibration
\cite{morone2020fibration} since it collapses the graph into its
minimal base capturing the minimal number of fibers (or balanced
colors) of the graph, and thus collecting the maximal symmetries of
the graph.

An implementation of the algorithm to find minimal balanced colorings
in a graph in the form of an R package is available at
\url{https://github.com/makselab/fibrationSymmetries} and
\url{https://osf.io/z793h/}.

\subsection{Fiber symmetries constrain admissible ODEs into synchronization}

The fibration symmetry of a graph imposes a structural constraint that
can facilitate synchronization. Specifically, if the graph has a
fibration symmetry, the nodes in each fiber can be expected to
synchronize with each other \cite{deville2015modular} due to the
consistent interaction patterns enforced by symmetry. This can be
understood through the synchronization of a system of ordinarily
differential equations (ODEs) 'admissible' to a graph
\cite{golubitsky2006nonlinear,boldi1999computing,avila2024fibration}.

Admissible means the structure of a graph $G$ imposes the coupling
terms between the $|V_G|$ number of ODEs, where each equation
characterizes the state of a vertex in the graph. Consider a dynamical
system on a graph where each vertex $v_i$ has a state $x_i$ that
evolves according to an ODE:
\begin{equation}\label{eq:ode}
\frac{dx_{i}}{dt} = F(x_{i},t) + \sum_{j \in \partial_i} A_{ij}H(x_{i},x_{j},t) .
\end{equation}

Here, $F$ represents the intrinsic dynamics of each vertex, $H$
embodies the interaction between vertices, and $A_{ij}$ are the
elements of the adjacency matrix $A$ of the graph, where $A_{ij}=1$
indicates the presence of an edge and $A_{ij}=0$ indicates the absence
of an edge from vertex $j$ to $i$. $\partial_i$ denotes the set of neighbors
of vertex $i$. As an example, the Kuramoto model, taken along with its
master stability function can be used to study synchronization in a
network
\cite{pecora2014cluster,kalloniatis2016fixed}.

\smallskip
Cluster synchronization as defined by Pecora {\it et al.}
\cite{pecora2014cluster} in this context refers to the situation in
which the dynamical states of a group of vertices for a given fiber
converge to a common trajectory, i.e., $x_{i}(t) \rightarrow
c_{\ell}(t) \; \forall \: v^{\ell}_{i}\in C_{\ell}$, given the
appropriate initial conditions and after any transients have died out.




The symmetries in the graph fibration allow the system of ODEs
admissible to the graph $G$ to be reduced from a $|V_G|$ number of
equations to a $K$ number of equations for each fiber
cluster. Equation \eqref{eq:ode} can be reduced to a system of equations
for each of the fibers:
\begin{equation}\label{eq:ode_fiber}
\frac{dc_{i}}{dt} = F(c_{i},t) + \sum_{j \in V(c_i)} Q_{ij}H(c_{i},c_{j},t).
\end{equation}

Here, $Q$ is the adjacency matrix of the base of the graph obtained by
the fibration.

The fibration formalism extends the automorphism symmetry groups of
the graph \cite{pecora2014cluster}. An automorphism is a permutation
symmetry of the nodes of the graph that leaves invariant the adjacency
of nodes. That is, the nodes permuted by the automorphism have the
same in- and out-neighbors before and after the application of the
automorphism. The analogous of fibers in fibrations are the orbits of
the automorphisms. An orbit a node is the set of nodes obtained by the
application of all the automorphisms of the graph.

Both orbits and fibers are balanced colorings of the graph. All orbits
are fibers, but not all fibers are orbits. Thus, the fibers of the
fibration capture more balanced colorings than the orbits of
automorphisms. In short, fibration symmetries are rigorous extensions
of automorphisms, and all automorphisms are fibration symmetries but
the opposite is not always true.

This situation is exemplified in Fig. \ref{fig:fibration}A. This graph
has no automorphism. That is, there is no permutation of any nodes
that leaves invariant the adjacency matrix. Therefore it has a trivial
orbital (and balanced coloring) partition where each node is its own
orbit or its own color. However, there is another minimal balanced
coloring partition that is shown in the two colors in the center of
Fig. \ref{fig:fibration}A. This is the balanced coloring partition
captured by the fibration, which is then applied to reduce the
network to the base on the left of Fig. \ref{fig:fibration}A.

A final fundamental difference between automorphism and fibration
symmetry is that the former is a global symmetry of the network. That
is, it is a permutation that constrains the global adjacency of nodes
to be the same. However, the fibration is a local symmetry leaving
invariant the input trees of the nodes, which are local views of the
network of the nodes collapsed by the fibration.  This fundamental
property makes the fibration applicable to a larger set of complex
networks than the more restricted automorphism.


\smallskip

The interplay between fibration symmetry and synchronization in
networks via admissible ODEs provide profound insights into the
dynamics of complex systems. The structural constraints imposed by
fibration symmetry can dictate the synchronization patterns, where the
quotation or base graph provides a simplified version of a larger
networks leading to predictable and potentially controllable behavior
in networks. This understanding has significant implications for the
design and analysis of complex systems in various domains, ranging
from biological networks (as in this paper) to engineered distributed
systems
\cite{pecora2014cluster,boldi1999computing,sorrentino2016complete,pournaki2019synchronization,morone2019symmetry,morone2020fibration,leifer2020circuits,leifer2021predicting,gili2024fibration}

\subsection{Symmetry Driven Repair Algorithm} \label{sec:mip}

In this section, we present a set of equations and inequalities based
on the concepts discussed previously to construct our integer linear
programming model for our problem, considering a specific graph, node
clustering, parameters, and constraints
\cite{ceria1998cutting,korte2018combinatorial}. We name this as the
Symmetry-Driven Repair Algorithm or SymRep for short.

\smallskip
For this model, we consider a directed graph where we denote $n=|V|$ and $m=|E|$ as the number of nodes and directed edges, respectively. We also define
\[
E^C = \{ (i,j) : i,j \in V, ij \not\in E\}
\]
as the set of node pairs between which no directed edge exists in $G$: these pairs indicate potential edges that could be added to the graph $G$. We define $\cS$ as a coloring of $G$, $\cS$ represents the sets dividing $V$, i.e., the various clusters of nodes.

The three types of decision variables in the model are:
\vspace{3mm}

\begin{tabular}{ll}
$r_{ij}$ for $(i,j) \in E$ & is a binary indicator for when edge
  $(i,j)$ is removed; \\ $a_{ij}$ for $(i,j) \in E^C$ & is a binary
  indicator for when a new edge $(i,j)$ is added; and \\ $s_{ijR}$ for
  $i \in P, j \in Q, R, P, Q \in \cS$ & is a binary indicator that $i$
  and $j$ has an imbalance in the color set $R$.
\end{tabular}

\vspace{3mm}

The objective function aims to minimize the weighted sum of edges
removed and added, defined as
\begin{equation}
\label{eq:mip_obj}
f_{\alpha,\beta}(r,a) = \alpha \sum_{ij \in E} r_{ij} + \beta \sum_{ij \in E^C} a_{ij}.
\end{equation}
Constants $\alpha, \beta$ are parameters that adjust the importance
between edge removal and addition in the objective.  The main
constraint ensures that $\cS$ represents a balanced coloring of the
graph $G$, as defined by

\begin{align}
    \nonumber
    \sum_{ip \in E: i \in S} (1 - r_{ip}) + \sum_{ip \in E^C:i \in S} a_{ip} & = \\
    \sum_{iq \in E: i \in S} (1 - r_{iq}) + \sum_{iq \in E^C: i \in S} a_{iq}; 
   & \quad p,q \in T; S,T \in \cS.
       \label{eq:balancing}
\end{align}

This aims to ensure the nodes in the same division receive equal
influence from all other divisions; this is the SymRep equivalent of
Eq. \eqref{eq:equality}. An optional constraint ensures that the
in-degree of all nodes is at least one.
\begin{equation}
    \label{eq:atleastone}
    \sum_{ip \in E} (1 - r_{ip}) + \sum_{ip \in E^C} a_{ip}
    \geq 1, \quad \quad \quad p \in V.
\end{equation}
This helps avoid scenarios where nodes in a fiber are disconnected
from the rest of the network. The following constraints are necessary
but not sufficient for minimally balanced colorings
Eq. \eqref{eq:unequality}.
\begin{equation}
\begin{split}
    \sum_{ip \in E: i \in R} (1 - r_{ip}) + \sum_{ip \in E^C :i \in R}
    a_{ip} - \left(\sum_{iq \in E: i \in R} (1 - r_{iq}) + \sum_{iq
      \in E^C :i \in R} a_{iq}\right) \geq s_{pqR} + ns_{qpR}; \\ p
    \in S; q \in T; R,S,T \in \cS
\end{split}
\label{eq:minimal_nec1}
\end{equation}
\begin{equation}
\begin{split}
    \sum_{iq \in E: i \in R} (1 - r_{iq}) + \sum_{iq \in E^C :i \in R}
    a_{iq} - \left(\sum_{ip \in E: i \in R} (1 - r_{ip}) + \sum_{ip
      \in E^C :i \in R} a_{ip}\right) \geq s_{qpR} + ns_{pqR}; \\ p
    \in S; q \in T; R,S,T \in \cS,
\end{split}
\label{eq:minimal_nec2}
\end{equation}
\begin{equation}
    s_{pqR} + s_{qpR} \leq 1; \quad p \in S; q \in T; R,S,T \in \cS,
\label{eq:minimal_nec3}
\end{equation}
\begin{subequations}
\begin{align}
    \sum_{R \in \cS} (s_{pqR} + s_{qpR}) &\geq 1; & p \in S; q \in T;
    S,T \in \cS \label{eq:minimal_nec4_A} \\ \sum_{O \in \cS-(S\cup
      T)} (s_{pqO} + s_{qpO}) + \sum_{I \in (S\cup T)} |s_{pqI} -
    s_{qpI}| &\geq 1; & p \in S; q \in T; S,T \in
    \cS \label{eq:minimal_nec4_B}
\end{align}
\label{eq:minimal_nec4}
\end{subequations}
These inequalities define the limits to ensure that two different
divisions do not have the same influence, termed the unbalancing
constraint. The entire model is defined as follows.
\begin{equation}
\label{eq:complete}
\min\left\{f_{\alpha,\beta}: \eqref{eq:balancing},
\eqref{eq:atleastone}, \eqref{eq:minimal_nec1},
\eqref{eq:minimal_nec2},\eqref{eq:minimal_nec3},
\eqref{eq:minimal_nec4}, r_{ij}, a_{k\ell}, s_{pqR} \in \{0,1\}, ij
\in E,k\ell \in E^C, p \in P, q \in Q, P\not=Q,R \in \cS\right\}.
\end{equation}
where Eq. \eqref{eq:minimal_nec4} within the equation above is a
reference to only select one of its
sub-equations. Eq. \eqref{eq:minimal_nec4_B} restricts feasible
solutions and potentially better minimal balanced solutions relative
to Eq. \eqref{eq:minimal_nec4_A} but is stronger with respect to the
minimal balanced coloring property. In our implementation, we always
try to find a solution to our networks by first implementing
Eq. \eqref{eq:minimal_nec4_B} after which if it fails to produce a
\textit{minimal} colored solution, then Eq. \eqref{eq:minimal_nec4_A}
is used.  We solve the integer linear programs with the solver
Gurobi. The implementation is available at
\url{https://github.com/makselab/PseudoBalancedColoring} and
\url{https://osf.io/26u3g}.

\section*{Acknowledgments}
We acknowledge discussions with Ian Leifer and Luis Alvarez. We thanks Ev Yemini for the NeuroPal animals and advice. 

\section*{Funding}
Funding was provided by NIBIB and NIMH through the NIH BRAIN Initiative Grant  R01 EB028157 and ONR Grant N00014-22-1-2835. The work in this manuscript was partially supported by the Simons Foundation (\#543069) (M.Z.) and the International Research Scholar Program by the Wellcome Trust and Howard Hughes Medical Institute (\#208565/A/17/Z) (M.Z.).

\section*{Authors contributions}
B.A. and T.G. performed the analysis of the data. P.A. performed Ca++ imaging experiments, neuron identifications, and the analyses shown in Fig. 2. D.P. performed the algorithmic developments. M.Z. led the experimental work and provided advice for the theory part. H.A.M. led the modeling and theoretical work. B.A., P.A., D.P., T.G., M.Z., and H.A.M. wrote the manuscript. M.Z. and H.A.M. led the study. 

\section*{Data and code availability}
All data and code are available at \url{https://github.com/MakseLab},
\url{https://osf.io} and \url{http://kcorebrain.com}.  In particular,
an implementation of the algorithm to find minimal balanced colorings
in a graph in the form of an R package is available at
\url{https://github.com/makselab/fibrationSymmetries} and
\url{https://osf.io/z793h/}. An implementation of the MILP solver is
available at \url{https://github.com/makselab/PseudoBalancedColoring}
and \url{https://osf.io/26u3g}.
    
\bibliography{bibliography.bib}
\bibliographystyle{naturemag}

\clearpage
\section*{Supporting information}

\begin{figure}[H]
\centering
\includegraphics[width=\linewidth]{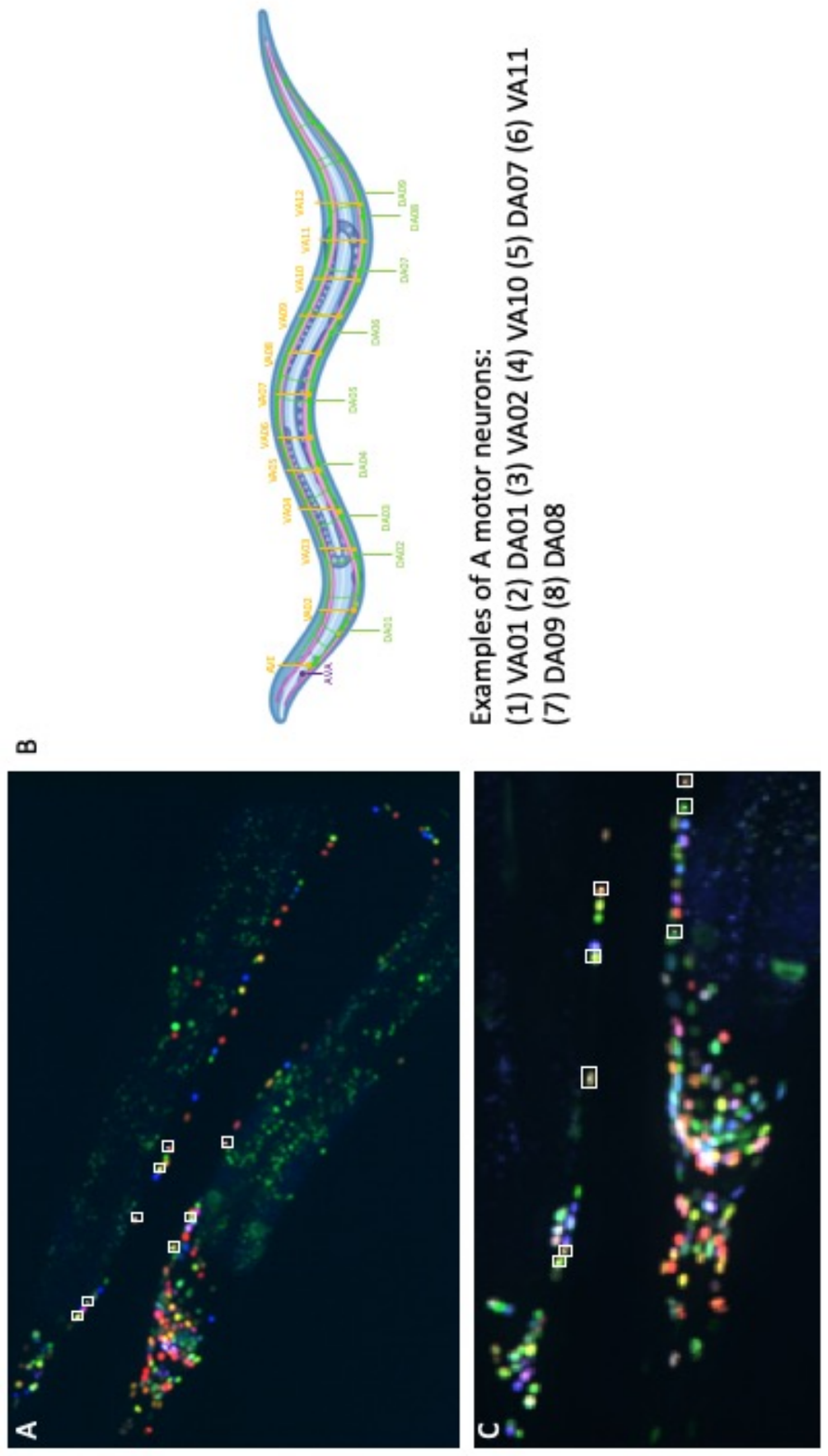}
\caption{{\bf Motor neuron examples through NeuroPAL labeling.}  (A)
  20x magnification high-quality NeuroPAL Z stack labeling with
  some specific examples of reverse motor neurons are also present in (C).
  (B) Schematic representing the anatomical location of all motor
  neurons of interest (DAs and VAs). (C) 40x magnification
  high-quality NeuroPAL Z stacks labeling reverse motor neurons also
  present in (A).}
\end{figure}

%
%
%

\end{document}